%% file: main.tex
\documentclass[12pt]{iopart}
\usepackage{iopams}  
\usepackage{graphicx}
\usepackage{dcolumn}
\usepackage{bm}
\usepackage[utf8]{inputenc}
\usepackage[T1]{fontenc}
\usepackage{mathptmx} 
\usepackage{tikz-feynman}
\usepackage{float}
\usepackage{longtable}
\usepackage{amsmath}

\usepackage{subfigure}

\newcommand{\stdepsilon}{\text{\usefont{OML}{cmr}{m}{n}\symbol{15}}}

\usepackage[percent]{overpic}

\begin{document}

\include{tikzpics}

\title[Renormalization group analysis of a model with self-organized criticality]{Renormalization group analysis of a continuous model with self-organized criticality: Effects of randomly moving environment}

\author{N.~V.~Antonov,$^{1, 2}$ P.~I.~Kakin,$^{1}$ 
N.~M.~Lebedev$^{2,*}$ and A.~Yu.~Luchin$^{1}$ }

\address{$^1$ Department of Physics, Saint Petersburg State University,
7/9 Universitetskaya Naberezhnaya, Saint Petersburg 199034, Russia \\
$^2$ N.~N.~Bogoliubov Laboratory of Theoretical Physics, Joint Institute for Nuclear Research, Dubna 141980, Moscow Region, Russia}

\ead{n.antonov@spbu.ru, p.kakin@spbu.ru, nikita.m.lebedev@gmail.com, luhsah@mail.ru}
\vspace{10pt}

\begin{abstract}
We study a strongly anisotropic self-organized critical system coupled to an isotropic random fluid environment. The former is described by a continuous (coarse-grained) model due to Hwa and Kardar. The latter is modeled by the Navier–-Stokes equation with a random stirring force of a rather general form that includes, in particular, the overall shaking of the system and
a non-local part with power-law spectrum $\sim k^{4-d-y}$ that describes, in the limiting case $y\to4$, a turbulent fluid.
The full problem of the two coupled stochastic equations is represented as a 
field theoretic model which is shown to be multiplicatively renormalizable and logarithmic at $d=4$. Due to the interplay between isotropic and anisotropic interactions, the corresponding renormalization group (RG) equations reveal a rich pattern of possible infrared (large scales, long times) regimes of asymptotic behaviour of various Green's functions. 
The attractors of the RG equations in the five-dimensional space of coupling parameters include a two-dimensional surface of Gaussian (free) fixed points, a single fixed point that corresponds to the plain advection by the turbulent fluid (the Hwa--Kardar self-interaction is irrelevant) and a one-dimensional curve of fixed points that corresponds to the case where the Hwa--Kardar nonlinearity and the uniform stirring are simultaneously relevant.    
The character of attractiveness is determined by the exponent $y$ and the dimension of space $d$; the most interesting case $d=3$ and $y\to 4$
is described by the single fixed point. The corresponding critical dimensions of the frequency and the basic fields are found exactly.
\\
\\
*Author to whom any correspondence should be addressed.
\end{abstract}


\section{\label{sec:level1} Introduction}

In this paper we study critical behaviour of a self-organized critical (SOC) system subjected to a moving random environment. The SOC system is described by the anisotropic continuous Hwa--Kardar (HK) model \cite{HK,HK2,HK1} while the medium is described by the isotropic Navier--Stokes (NS) equation with a random stirring force of a very general form that covers, in particular, the overall shaking of the system and, in some limiting case, turbulent motion \cite{FNS,DM79,AVP83,UFN,RedBook}.

The very popular concept of SOC was initially based on specially designed discrete models; now it is considered one of the strongest contenders for the mechanism behind emergent complexity in Nature \cite{Bak,Bak1,Bak2,Bak3,Col3,Col2,Watkins,Complex}. To be precise, it might explain self-similar spatio-temporal correlations in the infrared (IR, large distances, long times) range believed to be observed in a startlingly big number of systems.
However, a firm evidence of SOC was found so far only in two discrete models (belonging to the same universality class) and in a handful of experiments \cite{Bak3,Watkins}. Thus, the very formulation of the concept of SOC and its relevance to the real World remain somewhat questionable. Nevertheless, the widening range of possible applications and numerous conceptual questions raised by the non-equilibrium critical behaviour still fuel interest in further studying SOC. 

Originally, the concept of critical state arose in the theory of equilibrium systems in the vicinity of their second-order phase transition points. Later on, fruitful and guiding ideas of critical scaling, universality classes, effective fluctuation Hamiltonians, renormalization group flows were formulated and elaborated within the framework of the field theory of critical state; see, {\it e.g.} \cite{Fisher50,Cimento}  and the references therein.

That experience had shown that it is quite natural to use continuous field theory instead of discrete models. In particular, the Ising and Heisenberg lattice models for ferromagnets can be substituted by the continuous $O(n)$-$\phi^4$ field theory as long as their critical behaviour is concerned; see, {\it e.g.} \cite{Zinn,Book3} and references therein.

The dynamics of such nearly-equilibrium systems can also be described by continuous stochastic Langevin-type equations; see Chap.~5 in \cite{Book3}. 
The general Martin--Siggia--Rose--De~Dominicis--Janssen approach (see~\cite{Zinn,Book3} for discussion and references) allows one to reformulate such stochastic equations as certain field theoretic models, which opens the possibility for application of well-developed advanced methods: functional integral representations, exact Schwinger--Dyson equations and Ward identities, diagrammatic perturbation theory, analytic and dimensional regularization, renormalization theory and especially the renormalization group (RG). If such a model happens to be multiplicatively renormalizable, or it can be generalized to be of this kind, the corresponding RG equations are derived in standard fashion, and the IR asymptotic behaviour is governed by IR attractive fixed points (or more general manifolds like sets of points, lines, surfaces or limiting cycles). Then the scaling behaviour of the relevant quantities (correlation, Green's or response functions) naturally follows, sometimes with regular perturbative expansions for the corresponding critical dimensions and universal scaling functions.

With time, those ideas were extended to numerous non-equilibrium systems and have been successfully applied there for more than fifty years; see  \cite{Fisher50,Cimento,Tauber,Tauber2,APS,Wiese22} and references therein.
Thus, it was quite natural to apply the field theoretic approach to the models demonstrating SOC \cite{HK,HK2,HK1}; see also \cite{Diaz-Guilera,Diaz-Guilera2,Volk1,Volk2}.

However, there is a serious conceptual difference: criticality in nearly-equilibrium and numerous strongly non-equilibrium systems is usually achieved by fine-tuning certain ``control'' parameters (like temperature) to their critical values. On the contrary, the systems with SOC are believed to either self-tune this parameter to its critical value or, in the cases where such parameter is not in evidence, to arrive at a critical state without any tuning taking place.

In the field theoretic RG, the critical point of the $O(n)$-$\phi^4$ is described by the unstable fixed point of the RG equations: it becomes IR attractive when the temperature is tuned ``by hands'' to its critical value. From the same point of view, SOC is demonstrated by any field theoretic model with an IR attractive fixed point.
However, in the SOC community, such models are disregarded 
as representing only "generic" scaling behaviour. From this point of view, SOC 
models cannot be substituted by continuous models without ``throwing  the baby out with the bathwater.''

In this respect, the HK treatment of the a ``running'' sandpile model 
is remarkable for two reasons. Firstly, it provides an attempt to construct a continuous stochastic equation ({\it i.e.} an effective coarse-grained, smoothed, large-scale description), the derivation of which is based solely on conservation laws, relevant symmetries and dimensionality considerations. Secondly, its main goal was to disclose and illustrate proposed mechanism behind SOC, namely, transport in a driven diffusive system with anisotropy and conservation.

Admittedly, the HK model does not include one of the important features of SOC, namely, separation of time scales, and, moreover, it is not clear whether such a continuous description can adequately differentiate between different discrete sandpile models. At the same time, later studies disproved some of the assumptions made about the conditions necessary for SOC; see, {\it e.g.} Sec 9.2 in \cite{Bak3} for more details. This is why the HK model is conceptually closer to continuous stochastic models of generic scaling like surface roughening \cite{Zhang} or landscape erosion \cite{Pastor1,Pastor2,AK1,Duclut}. 

It is well known that internal disturbances such as static disordered impurities \cite{Prudnikov} or external effects of randomly moving environment~\cite{Mobilis,Onuki2,Nelson,Satten,Nandy,AHH} can drastically change the behaviour of a nearly-critical system. In particular, the HK model in the presence of quenched random drift velocity was considered in \cite{Disorder}.

Thus, more complete understanding of the situation can be achieved if one considers a SOC system interacting with its environment.
  In the series of works \cite{AK19,Serov1,Serov2,WeU,WeU2,HKNS1,HKNS2,HKNS3}, we studied the influence of environment motion on the SOC system described by the HK model. 
 In the earlier papers, we used ``synthetic'' Gaussian ensembles 
 due to Kraichnan and Avellaneda--Majda and their generalizations (general spatial dimension $d$ and finite correlation time) \cite{AK19,Serov1,Serov2,WeU,WeU2}.
 For more generality, it is interesting to take some dynamical stochastic  nonlinear process for the environment, for example, a random or turbulent fluid described by a stirred NS equation \cite{HKNS1,HKNS2,HKNS3}.

 The obtained results revealed complex patterns of possible critical behaviour including: the lack of crossover regime \cite{AK19}, non-universality and strong dependence on the type of random noise in the HK equation \cite{Serov1,Serov2} and non-conventional anisotropic scaling behaviour with a kind of dimensional transmutation \cite{WeU,WeU2}. When the motion was described by the version of the NS equation that models overall shaking of a fluid container, the leading-order RG analysis revealed a whole curve of simultaneously stable fixed points with an end point corresponding to the regime of the ``pure'' advection \cite{HKNS1,HKNS2,HKNS3}.
  
 The aim of the present paper is to go further and consider a more complex form of the NS equation that allows to model several types of random motion that include stirring, shaking and turbulent advection.

The plan of the paper is as follows. Section~\ref{sec:levelM} presents detailed discussion of the derivation of the HK model (Section~\ref{sec:levelM1}) and inclusion of the motion of the environment (Section~\ref{sec:levelM2}). The important role is played by the conservation law and the Galilean symmetry. Approximations made and possible generalizations are discussed. In Section~\ref{sec:level3}, 
we introduce the field theoretic formulation of the stochastic model. The ultraviolet (UV) divergences in the Green's functions are analyzed. Using the dimensional considerations, conservation and symmetry, the multiplicative renormalizability of the model is established. The corresponding renormalization constants are presented in the leading one-loop order of the perturbation theory. In Section~\ref{sec:levelA}, we derive the RG equations, define the corresponding RG functions ($\beta$ functions and anomalous dimensions $\gamma$) and present their explicit one-loop expressions. Then we discuss the attractors of the RG equations. In Section~\ref{sec:CD} we present {\it exact} expressions for the corresponding critical dimensions. In Section~\ref{idea} we discuss possible effect of the two-loop corrections. We argue that they may result in appearance of additional, fully nontrivial fixed point, where both the HK nonlinearity and turbulent advection are simultaneously important. Section~\ref{sec:Conc} is reserved for discussion and conclusion, while some relevant technical details are presented in the Appendix.

\section{\label{sec:levelM} Description of the model}

\subsection{\label{sec:levelM1} The Hwa--Kardar model}

The running sandpile model with partly open boundary conditions is a canonical model in SOC; see~\cite{HK,HK2,HK1}.
Being of rather general importance, the model is intuitively best illustrated by ``sand in a box with an open top and an open side.'' 
However, it can also be applied, {\it e.g.} for description of turbulent plasma transport in magnetic confinement devices \cite{Diamond95,Newman96,Carreras02}.
Originally, the evolution  is discreet in time and space and restricted by 
the box's boundaries.
The system is manifestly anisotropic: the new ``sand'' entering from above drives avalanches that cause some sand to exit through the side. While the sand pile surface is considered to be flat on average, locally it is getting rougher with time. The surface tilt is specified by a constant unit vector ${\bf n}$ with $|{\bf n}|=1$.

The continuous version of the HK model~\cite{HK,HK2,HK1} was designed by applying the ideas of hydrodynamic approach to the running sandpile model dynamics.
The basic coarse-grained (continuous) field 
$h=h(x)$ with $x =\{t, {\bf x}\}$ is the deviation of the local height of the pile profile over the $d$-dimensional substrate from its mean value. It satisfies a stochastic differential equation of motion that must incorporate the basic characteristic features of the initial sandpile model, namely:

1) {\it local conservation of the amount of particles by the bulk dynamics;} 

2) {\it addition of particles at random times in random positions by a driving force as well as particles dissipation at the open (free) boundaries} of the system;

3) {\it anisotropic nature of the mass transport in the bulk inherited from the boundary conditions of the pile; }

4) {\it the symmetry with respect to the combined reflection $x_{\parallel} \to -x_{\parallel};\ h \to -h$, replacing hills with pits simultaneously with a change in the preferred direction of their movement.}

The first two features can be implemented by choosing the equation of motion for the field $h(x)$ in the form of continuity equation with additional stochastic noise term modeling both the driving force and dissipation at the boundaries:
\begin{equation}
    \partial_{t} h= \boldsymbol{\partial}\, {\bf j} + \eta = \partial_i j_i + \eta
    \label{continuity}
\end{equation}
with a certain current ${\bf j} = \{ j_i(x) \}$.
In equation (\ref{continuity}) and below, $\boldsymbol{\partial}=\{\partial_i = \partial / \partial x_i\}$
and summation over repeated indices $i=1,\dots,d$ is implied.
The noise $\eta=\eta(x)$ is also meant to model interaction with otherwise discarded microscopic degrees of freedom. Thus, it is ``external'' in the sense of \cite{Diaz-Guilera, Diaz-Guilera2} and therefore does not respect the conservation.

In the formulation of \cite{HK,HK2,HK1}, which we also naturally take to apply in our study, the infinite volume with no boundaries is considered, while the original anisotropic set up survives in the single unit vector ${\bf n}$ that enters the basic equations as a signature of the preferred direction of the mass transport in the $d$-dimensional space.
Thus, any vector can be decomposed in the components transverse and parallel to that direction, in particular,
\[
{\bf x} = {\bf x_{\bot}} + {\bf n} x_{\parallel}, \quad 
x_{\parallel} =({\bf n}{\bf x}), \quad ({\bf n}{\bf x_{\bot}})=0
\]
for the coordinate and 
\[
\boldsymbol{\partial}=\boldsymbol{\partial}_{\bot} + {\bf n} \partial_{\parallel},
\quad \partial_{\parallel} = ({\bf n}\boldsymbol{\partial}),  \quad ({\bf n}\boldsymbol{\partial}_{\bot})=0
\]
for the gradient, so that
\[
\boldsymbol{\partial}_{\bot} =\{\partial_i^{\bot}= \partial_i - n_i (n_k\partial_k)\} ,\quad
(\boldsymbol{\partial}\boldsymbol{\partial}_{\bot}) =
(\boldsymbol{\partial}_{\bot}\boldsymbol{\partial}_{\bot})\equiv 
\partial^2_{\bot}.
\]

The current ${\bf j}$ in (\ref{continuity}) has to be constructed out of the two available vectors ${\bf \partial}$ and ${\bf n}$, with certain scalar coefficients. As one is interested in the IR behaviour, only leading-order terms in the gradient expansion should be kept. Moreover, it is assumed that fluctuations of the surface profile are not overly large, so that only terms linear and quadratic in $h$ are included. All these considerations are typical of the theory of critical behaviour and can be justified {\it a posteriori} by the analysis of canonical dimensions; see, {\it e.g.} Section~1.16 in~\cite{Book3}. Along with the symmetry  $x_{\parallel} \to -x_{\parallel};\ h \to -h$, this gives:

\begin{equation}
    {\bf j}_{HK} = -\nu_{\bot}\boldsymbol{\partial}_{\bot} h - \nu_{\parallel} 
    \partial_{\parallel} h\, {\bf n}  +  \lambda  h^{2} {\bf n}.
    \label{current}
\end{equation}

Substitution into (\ref{continuity}) results in:
\begin{equation}
\partial_{t} h= \nu_{\bot}\, \partial_{\bot}^{2} h + \nu_{\parallel}\, \partial_{\parallel}^{2} h - 
\partial_{\parallel} h^{2}/2 + \eta.
\label{HK}
\end{equation}
Following Hwa and Kardar we define the Gaussian random noise $\eta(x)$ in (\ref{HK}) by its zero mean and prescribed pair correlation function:
\begin{equation}
\langle \eta(x)\eta(x') \rangle = C\,
\delta(t-t')\, \delta^{(d)}({\bf x}-{\bf x}'), \quad
\label{forceC}
\end{equation}
where $C>0$ is a measure of the noise intensity. Although an additional internal noise term that obeys conservation can also be introduced, it will appear to be IR irrelevant and hence can be dropped from the outset \cite{Diaz-Guilera, Diaz-Guilera2}.

\subsection{\label{sec:levelM2} Inclusion of the environment}

Consider how the interaction with a moving environment, described by the vector velocity field ${\bf v}(x)=\{v_i(x)\}$ with $i=1,\dots, d$, can be included into the current ${\bf j}$ from equation  (\ref{continuity}) in the spirit of the HK phenomenology. We take:
\begin{equation}
    {\bf j} = {\bf j}_{HK} + 
    \alpha_{is}\,  h\, {\bf v} + 
    \alpha_{an}\, ({\bf n v})\,h\,{\bf n} + 
    \beta_{is}\, v^2\,{\bf n} + 
    \beta_{an}\, ({\bf vn})^2\,{\bf n},
    \label{current_long}
\end{equation}
where ${\bf j}_{HK}$ is the current (\ref{current}) and $\alpha$,  $\beta$ are certain constant scalar coefficients. In this expression, we included only terms quadratic in the set of fields and discarded IR irrelevant terms that contain derivatives and extra powers of the fields. 

Let us skim through the new terms in (\ref{current_long}). The $\beta$ terms explicitly violate Galilean symmetry expected to hold in the bulk (at least for the large enough system, which we implicitly assume). Thus, for now, we omit these terms, reserving the study of their effect for the upcoming research.

The $\alpha_{an}$ term in (\ref{current_long}) represents the advection of the field $h$ strictly along the preferred direction ${\bf n}$ and,
therefore, introduces anisotropy into the advection process.
The effects of such strongly anisotropic advection, represented solely by this term, were studied in our previous works \cite{AK19, Serov1, Serov2}. There, two different anisotropic synthetic ensembles were used to model the statistics of the one-component
velocity field ${\bf v} = {\bf n}\,v(t, {\bf x_{\bot}})$. 

Inclusion of both the $\alpha$ terms allows one to study the interplay between isotropic and anisotropic modes of advection. However, 
here we are interested mostly in the interaction of strongly anisotropic mass transport in the original HK model with purely isotropic advection. In particular, we will see that subtle differences in the statistics of the isotropic advecting field ${\bf v}$
may lead to remarkable changes in the resulting IR behaviour. Thus, for now, we omit the $\alpha_{an}$ term, leaving the study of its effects for the future.

After all said above, what we are left with is the only term proportional to $\alpha_{is}$, which represents fully isotropic advection. Being again substituted to the continuity equation, this augmented current produces two new terms in the equation (\ref{HK}):
\begin{equation}
\partial_{t} h= \nu_{\bot}\, \partial_{\bot}^{2} h + \nu_{\parallel}\, \partial_{\parallel}^{2} h - \partial_{\parallel} h^{2}/2 + \alpha_{is}\, {\bf v} \boldsymbol{\partial} h + \alpha_{is}\, h\, \boldsymbol{\partial} {\bf v} + \eta.
\label{HKadjusted}
\end{equation}
Finally, we will restrict ourselves to the case of 
incompressible fluid, so that the field ${\bf v}$ is 
transverse: ${\bf  \partial} {\bf v} = 0$ or, in the momentum representation, ${\bf  k} {\bf v} = 0$. Then the second new term vanishes and we are left with the only term proportional to $\alpha_{is}$. In turn, this coefficient can now be absorbed by the rescaling of the field ${\bf v}$; for convenience, we will fix $\alpha_{is} = -1$. We arrive at the final equation:
\begin{equation}
\nabla_{t} h= \nu_{\bot}\, \partial_{\bot}^{2} h + \nu_{\parallel}\, \partial_{\parallel}^{2} h - \partial_{\parallel} h^{2}/2 + \eta,
\label{HKadjusted-fin}  
\end{equation}
where
\begin{equation}
\nabla_{t} = \partial_{t} + ({\bf v} {\bf \partial}) =  \partial_{t} + v_k \partial_k
\label{nabla-fin}
\end{equation}
is the Lagrangian (Galilean covariant) derivative. It is natural to interpret this equation as an advection of the bulk material by the movement of the encompassing environment.

The motion of the medium will be described by the stochastic differential NS equation for an isotropic incompressible viscous fluid:
\begin{equation}
\nabla_t v_{i} = \nu \partial^2 v_{i} - \partial_i \wp + f_i,
\label{NS}
\end{equation}
where $\wp(x)$ is the pressure, $\nu$ is the kinematic viscosity coefficient, and  ${\bf f}(x)=\{f_i(x)\}$ is an external random transverse stirring force.

The common choice for the statistics of the force is the Gaussian distribution with a zero mean and a prescribed correlation function:
\begin{equation}
\langle f_i (t, {\bf x}) f_j (t',{\bf x}') \rangle= 
\delta(t-t')\, D_{ij} ({\bf x}-{\bf x}').
\label{forceD}
\end{equation}
The temporal $\delta$ correlation in (\ref{forceD}) is required to preserve the Galilean symmetry of the equation (\ref{NS}). The requirement of the transversality of the vector field ${\bf v}$ translates to the bond equation for the pressure $\wp = - \partial^{-2}\partial_i\partial_k v_i v_k -  \partial^{-2}\partial_{i}\eta_{i}$. This choice of pressure simultaneously ensures the transversality of the velocity field and completely absorbs the longitudinal part of the random force. Hence, without loss of generality, one can choose ${\bf \eta}$ to be divergence-free, so that its pair correlation function is proportional to the transverse projector $P_{ij}({\bf k}) = \delta_{ij} - k_i k_j / k^2$:
\begin{eqnarray} 
D_{ij} ({\bf x}-{\bf x}')= \, \int \frac{d\mathbf{k}}{(2\pi)^d}
P_{ij} (\mathbf{k})\, D (k)    \: \mathrm{exp} \,\{ i(\mathbf{k}(\mathbf{x-x'}))\},
\label{forceD1}
\end{eqnarray}
where 
$k\equiv |{\bf k}|$ is the wave number. We choose the function $D(k)$ that includes two terms
\begin{equation}
D (k) = D_1 + D_2 k^{4-d-y}, 
\label{loc-nonloc}
\end{equation}
where $D_1$ and $D_2$ are positive amplitude factors. 

Strictly speaking, it is not necessary to take the force $f_i$ to be transverse; its longitudinal part will anyway be ``swallowed'' by the pressure term $\wp$ in the NS equation (\ref{NS}). Thus, the transverse projector in the correlation function  (\ref{forceD1}) is in fact superfluous: it will effectively be restored within the diagrams of the perturbation theory. Once it is omitted, we can interpret the first term in (\ref{loc-nonloc}) as a constant or uniform stirring in the $k$ representation; in the coordinate space it is local as it corresponds to $\delta({\bf x}-{\bf x'})$. Moreover, its $k=0$ component can be interpreted as an overall shaking of the fluid container~\cite{FNS}.

On the contrary, the $D_2$ term in (\ref{loc-nonloc}) is neither local nor uniform: it depends on $k$ and in the coordinate representation describes a long-range correlation of the form $\propto |{\bf x}-{\bf x'}|^{-4+y}$, often used to model fully developed turbulence \cite{DM79,AVP83}; see also \cite{UFN,RedBook,Book3}. Then the most interesting value of the exponent $y$ is provided by the limit $y\to4$, when the function (\ref{forceD}) with $D_1=0$ and $D_2 \sim (4-y)$ can be viewed as a power-like representation of the function $\delta({\bf k})$, which models the energy input from the largest scales; see Section~2.8 in~\cite{RedBook} or Section~6.11 in~\cite{Book3}.

\section{\label{sec:level3}Field theoretic formulation, canonical dimensions, UV divergences and renormalization}

Using the Martin--Siggia--Rose--De~Dominicis--Janssen formalism (see, {\it e.g.} Section~5.3 in~\cite{Book3} and references therein) the full stochastic problem~(\ref{HKadjusted-fin})--(\ref{loc-nonloc}) is represented as a field theory that involves doubled set of fields $\Phi=\{h,h',{\bf v},{\bf v'}\}$ and the action functional
\begin{eqnarray}
    S(\Phi) = C_0 h'h'/2 + h'\{-\nabla_t h + \nu_{\parallel 0} \partial_{\parallel}^{2} h + \nu_{\bot 0} \partial_{\bot}^{2} h - 
\partial_{\parallel} h^{2}/2\} + S_v({\bf v},{\bf v'}),
\label{action_ve}
\end{eqnarray}
where
\begin{eqnarray}
S_v({\bf v},{\bf v'}) =  v'_i D_{ij} v'_j /2 +
v'_i\{-\nabla_t{v_i}+ \nu_0 \partial^{2} {v_i}\}.
\label{action_ve2}
\end{eqnarray}
Here $\nabla_t$ is from (\ref{nabla-fin}) and $D_{ij}$ is from (\ref{forceD1}); $h'$ and ${\bf v'}$ are so-called auxiliary fields which introduction is required by the formalism. For brevity, all the needed integrations over the arguments ({\it e.g.} such as $x=\{t,{\bf x}\}$ and $x'=\{t',{\bf x'}\}$ in the first term of the action (\ref{action_ve2})) are implied here and below in similar expressions.

The pressure term ${\partial}_{i}\wp$ is absent in the action~(\ref{action_ve2}) because it was integrated out (as the field ${\bf v}$ is transverse, so is the auxiliary field ${\bf v}'$) with the field ${\bf v}'$ now playing its role of transverse projector.

In order to identify ultraviolet (UV) divergences in the Green's functions of the theory (\ref{action_ve}), (\ref{action_ve2}), we use canonical dimensions analysis (see, {\it e.g.}~\cite{Book3}, Sections~1.15 and 1.16). Canonical dimension of any quantity $F$ is defined as (see~\cite{Book3}, Sections~1.17 and 5.14)
\begin{eqnarray}
[F] \sim [T]^{-d_{F}^{\omega}} [L]^{-d_{F}^{k}},
 \label{canonical1}
\end{eqnarray}
where $T$ is the time scale, $L$ is the length scale, $d_{F}^{\omega}$ is the frequency dimension, and $d_{F}^{k}$ is the momentum dimension.

By taking into account normalization conditions
\begin{eqnarray}
d_{k_i}^k=-d_{x_i}^k=1,\ d_{k_i}^{\omega} =d_{x_i}^{\omega }=0,\
d_{\omega }^k=d_t^k=0,\  d_{\omega }^{\omega }=-d_t^{\omega }=1,
\label{normal_conditions}
\end{eqnarray}
one arrives at the results presented in Table~\ref{canonical dimensions} where we introduced a number of new notations: total canonical dimension $d_{F}$ is defined as $d_{F}=d_{F}^{k}+2d_{F}^{\omega}$ (see \cite{Book3}, Section 5.14); $\mu$ is the renormalization momentum scale (to appear below) and $\epsilon= 4-d$ measures deviation from logarithmic dimension.

\begin{table}[h]
\caption{Canonical dimensions 
in the theory~(\ref{action_ve}), (\ref{action_ve2}); $\epsilon= 4-d$.}
\label{canonical dimensions}
\begin{tabular}{|c||c|c|c|c|c|c|c|c|c|c|c|}
 \hline
$F$&$h(x)$&$h'(x)$&${\bf v}(x)$&${\bf v'}(x)$&$C_0$,\ $D_{10}$& $D_{20}$ &$\nu_0$,\ $\nu_{\parallel0}$,\ $\nu_{\bot0}$&$g_{0}$,\ $w_0$&$u_0$&\mbox{$x_{10}$,\ $x_{20}$}&{$\mu$}\\
 \hline
$d^{\omega}_F$&$1$&$-1$&$1$&$-1$&$3$&$3$&$1$&$0$&$0$&$0$&$0$\\
 \hline
$d^{k}_F$&$-1$&$d+1$&$-1$&$d+1$&$-d-2$&$y-6$&$-2$&$\epsilon$&$y$&$0$&$1$\\
 \hline
$d_F$&$1$&$d-1$&$1$&$d-1$&$\epsilon$&$y$&$0$&$\epsilon$&$y$&$0$&$1$\\
 \hline
\end{tabular}
\end{table}

The coupling constants $g_0$, $w_0$ and $u_0$, defined by the relations
\begin{eqnarray}
 g_0=C_0/\nu_{\bot0}^{3/2}\nu_{\parallel0}^{3/2},
 \quad w_0=D_{10}/\nu_{0}^3, \quad u_0=D_{20}/\nu_{0}^3,
 \label{couplings1}
\end{eqnarray}
become dimensionless at $\varepsilon=0$ and $y=0$, which means 
that the full model becomes logarithmic at $\varepsilon=y=0$ and that the
UV divergences manifest themselves as singularities in $\varepsilon$, $y$ and their combinations. Although the ratios 
\begin{eqnarray}
 x_{1,0} = \nu_{\parallel 0}/\nu_0, \quad x_{2,0} = \nu_{\perp 0}/\nu_0    \label{couplings2}
\end{eqnarray}
do not appear as expansion parameters in the perturbation theory, they are dimensionless and should be treated on an equal footing with the couplings (\ref{couplings1}) in the RG analysis.

The 1-irreducible Green's function $\Gamma$ can involve superficial UV divergence when its formal divergence index $\delta_{\Gamma}$, equal to its total canonical dimension $d_{\Gamma}$ taken at logarithmicity, 
\begin{eqnarray}
d_{\Gamma}&=&d+2-d_h N_h-d_{h'} N_{h'}-d_v N_{{\bf v}}-d_{{\bf v'}} N_{{\bf v'}},
\\
\delta_{\Gamma} &=& d_{\Gamma}|_{\varepsilon=y=0}=
6- N_h-3 N_{h'}- N_{{\bf v}}-3 N_{{\bf v'}}
\label{10cc}
\end{eqnarray}
is a nonnegative integer. However, the specific form of the interactions (a spatial derivative can be moved onto auxiliary fields using integration by parts)
reduces (\ref{10cc}) to the real divergence index
\begin{equation}
\delta_{\Gamma}'= \delta_{\Gamma} - N_{h'} - N_{{\bf v'}} =
6- N_h-4 N_{h'}- N_{{\bf v}}-4 N_{{\bf v'}}.
\end{equation}
Finally, the Galilean symmetry of the full model (\ref{action_ve}), (\ref{action_ve2})
\begin{eqnarray}
{\bf v}(t, {\bf x}) \to {\bf v}(t, {\bf x} +{\bf u}\,t) - {\bf u}, \quad 
{\bf v'}(t, {\bf x}) \to {\bf v'}(t, {\bf x} +{\bf u}\,t),
\nonumber \\
    h(t,{\bf x}) \to h(t,{\bf x}+{\bf u}\,t), \quad
h'(t,{\bf x}) \to h'(t,{\bf x}+{\bf u}\,t)  , \quad {\bf u}= {\rm const}
\label{galilean}
\end{eqnarray}
further reduces the form of the counterterms.\footnote{More detailed discussion can be found in Section~4 in our paper \cite{HKNS1}; the analysis is very similar to the present case.}

Thus, there are only three divergent 1-irreducible Green's functions left: $\langle h'h \rangle$, $\langle h'hh \rangle$ and $\langle {\bf v'}{\bf v} \rangle_{{\rm{1-ir}}}$, and the corresponding counterterms reduce to the forms:
\[
h'\partial_{\parallel}^2 h, \quad h'\partial_{\bot}^2 h, \quad h'\partial_{\parallel }h^{2}, \quad {\bf v'} \partial^2 {\bf v}.
\]
All such terms are already present in the action functional 
(\ref{action_ve})--(\ref{action_ve2}), so that the full stochastic model is multiplicatively renormalizable with the renormalized action functional:
\begin{eqnarray}
S_R(\Phi)& = g\mu^{\epsilon}\nu_{\bot}^{3/2}\nu_{\parallel}^{3/2} h'h'/2  + h'\{-\nabla_th+ Z_1\nu_{\parallel} \partial_{\parallel}^{2} h +
Z_2\nu_{\bot} \partial_{\bot}^{2} h - Z_{4} 
\partial_{\parallel} h^{2}/2\}+ \nonumber \\
& + w\mu^{\epsilon}\nu^3{\bf v'}^2/2+ u\mu^{y}\nu^3{\bf v'}^2/2+ {\bf v'}\{-\nabla_t {\bf v}+ Z_3\nu \partial^{2} {\bf v}\}.
\label{act_r}
\end{eqnarray}
The functionals (\ref{action_ve})--(\ref{action_ve2}) and (\ref{act_r}) are related by rescaling the fields $S_R(\Phi)=S(Z_{\Phi}\Phi)$ and by the following relations between the original (``bare'') parameters and their renormalized counterparts (without the subscript ``0''):
\begin{eqnarray}
g_0=Z_g g\mu^{\epsilon}, \quad w_0=Z_w w\mu^{\epsilon}, \quad u_0=Z_u u\mu^{y}, \quad x_{1,0}=Z_{x_1} x_1,\nonumber \\ x_{2,0}=Z_{x_2} x_2, \quad
    \nu_{0} = \nu_{} Z_{\nu} , \quad\nu_{\parallel0} = \nu_{\parallel} Z_{\nu_{\parallel}}, \quad \nu_{\perp0} = \nu_{\perp} Z_{\nu_{\perp}}.
    \label{renorm_param}
\end{eqnarray}
Here $\mu$ is the renormalization momentum scale (additional arbitrary parameter of the renormalized theory), introduced such that the renormalized coupling constants $g$, $w$ and $u$ are dimensionless with respect to the frequency and momentum canonical dimensions separately. The renormalization constants $Z_{f}$ from (\ref{renorm_param}) are related to those in (\ref{act_r}) as follows:
\begin{eqnarray}
    Z_{h} = Z^{-1}_{h'}=Z_{4}, \quad Z_{{\bf v}}=Z_{{\bf v'}}=1,
    \nonumber \\ 
    Z_g = Z_{1}^{-3/2}Z_{2}^{-3/2}Z_4^2, \quad Z_w = Z_u = Z_{3}^{-3}, \quad Z_{x_1} = Z_{1}Z_{3}^{-1}, \quad Z_{x_2} = Z_{2}Z_{3}^{-1}, 
    \nonumber \\
    Z_{\nu_{\parallel}}=Z_1, \quad Z_{\nu_{\perp}}=Z_2, \quad  Z_{\nu}=Z_3.
\label{ZZZ}
\end{eqnarray}
In the one-loop approximation, the renormalization constants in the minimal subtraction (MS) scheme have the forms:
\begin{eqnarray}
    Z_1 = 1-\frac{1}{\epsilon }\left[ g\frac{3}{16}+w\, f_1(x_1,x_2) \right] -\frac{u}{y} f_1(x_1,x_2), \label{Z1}\\ 
    Z_2 = 1 - \frac{w}{ \epsilon} f_2(x_1,x_2) -\frac{u}{y} f_2(x_1,x_2),
 \label{Z2}\\  
Z_3 = 1 - \frac{1}{ \epsilon} \frac{w}{8} - \frac{1}{ y} \frac{u}{8}, \label{Z3}\\
Z_4=1,
\label{ZZZZ}
 \end{eqnarray}
with possible higher-order corrections in $g$, $w$ and $u$. Here
\begin{eqnarray}
   f_1(x_1,x_2) =\frac{1}{ 2 x_1\left(\sqrt{1 + x_1} + \sqrt{1 + x_{2}} \right)^2}\left( 1 + 2\sqrt{\frac{1 + x_1}{1 + x_{2}}} \right);
   \label{ff}   \\
   f_2(x_1,x_2) =\frac{1}{ 6 x_2 \left(\sqrt{1 + x_1} + \sqrt{1 + x_{2}} \right)^2} 
    \left( 5 + 4\sqrt{\frac{1 + x_{1}}{1 + x_{2}}} \right) .
    \label{fff}
\end{eqnarray}
Some comments on the technique and details of the calculation are given in the Appendix.
It is not clear yet whether the last relation $Z_4=1$ in (\ref{ZZZZ})
is an exact fact or just an artefact of the one-loop calculation; this issue is discussed in Section~\ref{idea}.

\section{\label{sec:levelA} RG functions and attractors of the RG equations}

In a renormalizable model, the basic differential RG equation for a certain multiplicatively renormalized quantity $F=Z_F F_R$, 
such as a correlation or a Green's function, is derived in a standard way and has the form:

\begin{equation}
   \left( {\cal D}_{RG}+\gamma_F \right)\, F_R = 0,
\end{equation}
see, {\it e.g.}, Section~1.24 in~\cite{Book3}.
In our model the explicit form of the differential operator has the form:
\begin{eqnarray}
    {\cal D}_{RG} = {\cal D}_{\mu} + \beta_{g}\partial_{g} +
\beta_{w}\partial_{w} + \beta_{u}\partial_{u} + \beta_{x_1}\partial_{x_1} + \beta_{x_2}\partial_{x_2} -
\gamma_{\nu}{\cal D}_{\nu}, \nonumber
\end{eqnarray}
where ${\cal D}_{f}=f\partial_f$ for any quantity $f$ (a field or a parameter) and the coefficient functions are anomalous dimensions and $\beta$ functions:
\begin{eqnarray}
\gamma_{f} = \widetilde{\cal D}_{\mu}\ln Z_{f}, \quad
\beta_r = \widetilde{\cal D}_{\mu}r
\label{XXX}
\end{eqnarray}
for any quantity $f$ and any coupling constant $r$
with the differential operator $\widetilde{\cal D}_{\mu}$ defined as ${\cal D}_{\mu}= \mu\partial_{\mu}$ 
at fixed bare parameters $e_0= \{g_0, w_0, x_{10}, x_{20}, \nu_0\}$:
\begin{equation}
\widetilde{\cal D}_{\mu} = \mu\partial_{\mu}|_{e_0}.
\end{equation}
The relations (\ref{ZZZ}) between the renormalization constants translate to the relations between the anomalous dimensions:
\begin{eqnarray}
    \gamma_{h} = -\gamma_{h'}=\gamma_{4}, \quad \gamma_{{\bf v}}=\gamma_{{\bf v'}}=0,
    \nonumber \\ \gamma_g = -3 \gamma_{1}/2-3\gamma_{2}/2+2\gamma_4, \quad \gamma_w = -3\gamma_{3}, 
    \nonumber \\
    \gamma_{x_1} = \gamma_{1}-\gamma_{3}, \quad \gamma_{x_2} = \gamma_{2}-\gamma_{3}, 
    \nonumber \\
    \gamma_{\nu_{\parallel}}=\gamma_1, \quad \gamma_{\nu_{\perp}}=\gamma_2, \quad  \gamma_{\nu}=\gamma_3,
\label{gamma}
\end{eqnarray}
whilst the $\beta$ functions can be expressed through the anomalous dimensions as follows:
\begin{eqnarray}
\beta_g = -g \left[\epsilon +\gamma_{g} \right] = -g \left[\epsilon - \frac{3}{2}\gamma_{1} - \frac{3}{2}\gamma_{2} + 2\gamma_{4}  \right],
\label{bg}
\end{eqnarray}
\begin{eqnarray}
\beta_w = -w \left[\epsilon + \gamma_w \right] = -w \left[\varepsilon - 3\gamma_{3} \right],
\label{bw}
\end{eqnarray}
\begin{eqnarray}
\beta_u = -u \left[y + \gamma_u \right] = -u \left[y - 3\gamma_{3} \right],
\label{bu}
\end{eqnarray}
\begin{eqnarray}
\beta_{x_1} =  -x_1 \gamma_{x_1} = -x_{1} \left[\gamma_{1} - \gamma_{3} \right],
\label{bx1}
\end{eqnarray}
\begin{eqnarray}
\beta_{x_2} = -x_2 \gamma_{x_2} = -x_2 \left[\gamma_{2} - \gamma_{3} \right].
\label{bx2}
\end{eqnarray}

The explicit one-loop approximations for $\gamma_1$--$\gamma_4$ are obtained from the expressions (\ref{ZZZZ}) for the renormalization constants and have the forms:
\begin{eqnarray}
    \gamma_1 =  g\frac{3}{16}+(w+u)\,f_1(x_1,x_2), \nonumber \\ 
    \gamma_2 = (w+u) \,f_2(x_1,x_2), \quad
    \gamma_3 =  \frac{(w+u)}{8}, \quad \gamma_{4} = 0. 
    \label{op}
\end{eqnarray}
The last expression follows from the one-loop result $Z_4=1$; see the comment below (\ref{ZZZZ}) and further discussion in Section~\ref{idea}.

Hence the one-loop approximations for the $\beta$ functions of the coupling constants $g$, $w$, $u$, $x_1$, and $x_2$ read
\begin{eqnarray}
    \beta_{g} = -g \left\{ \epsilon - \frac{9}{32}\,g - \frac{3}{2}(w+u)\,\left[f_1(x_1,x_2)+f_2(x_1,x_2)\right] \right\},
    \label{b1}\\
    \beta_{w} = -w\left[ \epsilon - \frac{3}{8} (u + w) \right],
    \label{b2}\\
    \beta_{u} = -u\left[ y - \frac{3}{8} (u + w) \right],
    \label{b5}\\
    \beta_{x_{1}} = x_{1} \left\{ -\frac{3}{16}g + (w+u)\left[ \frac{1}{8} - \,f_1(x_1,x_2)\right] \right\},
    \label{b3}\\
    \beta_{x_{2}} = x_{2}(w+u)\left[  \frac{1}{8} - \,f_2(x_1,x_2) \right].
    \label{b4}
\end{eqnarray}

The roots of the system of $\beta$ functions $g^*$: $\beta(g^*)=0$, are coordinates of the fixed points of the RG equation. If the real parts of the eigenvalues $\lambda_i$ of the matrix $\Omega_{rn}=\{\partial\beta_{g_r}/\partial g_n\}|_{g^*}$ are positive for a given fixed point, that point corresponds to IR asymptotic behaviour; see, {\it e.g.}~\cite{Book3}, Section~1.42.

The following fixed points are found for the system (\ref{b1})--(\ref{b4}):

FP1: a surface of Gaussian fixed points:
\begin{eqnarray}
    g^{*} = 0; \quad w^{*} = 0; \quad u^{*} = 0; \quad x_{1}^{*}\neq 0; \quad x_{2}^{*} \neq 0;\nonumber\\
    \lambda_i=\{0,0,-y,-\epsilon,-\epsilon\}.
    \label{FP1}
\end{eqnarray}

FP2: a curve of fixed points: 

\begin{eqnarray}
    g^{*} = \frac{16\epsilon}{9}\left(1-8f_1(x_1^{*},x_2^{*})\right);\quad
    w^{*} = \frac{8}{3}\epsilon; \quad u^{*}=0 \quad
   f_2(x_1^{*},x_2^{*}) =\frac{1}{8};\nonumber\\
    \lambda_i=\{0,\varepsilon - y,\epsilon,\lambda_3,\lambda_4\}.
    \label{FP2}
\end{eqnarray}
The eigenvalues $\lambda_3$ and $\lambda_4$ are proportional to $\varepsilon$ for every point on the curve. 

The curve coincides with the one found in \cite{HKNS1} (see expression (53) there) and describes crossover regime of critical behaviour where both nonlinearity of the HK equation and local term in the correlator (\ref{forceD}), (\ref{forceD1}) are relevant simultaneously. Note that the power-like term is, on the contrary, irrelevant. Moreover, in the special case $y = \varepsilon$, charges $w$ and $u$ merge into the single charge $w+u$
as both terms in (\ref{forceD}), (\ref{forceD1}) become identical: $k^{\varepsilon-y} \sim k^{0}$. The eigenvalues for this special regime are $\lambda_i=\{0,0,\epsilon,\lambda_3,\lambda_4\}$.

FP2a: a point on the curve FP2 at which the HK nonlinearity becomes irrelevant:
\begin{eqnarray}
    g^{*} = 0; \quad w^{*} = \frac{8}{3}\epsilon; \quad u^{*}=0 \quad x_{1}^{*}=x_{2}^{*} = \frac{1}{2}(\sqrt{13} - 1);\nonumber\\
    \lambda_i=\left\{0, \varepsilon - y,\epsilon, \frac{47 + \sqrt{13}}{162} \epsilon, \frac{13 - \sqrt{13}}{18} \epsilon \right\}.
    \label{FP2a}
\end{eqnarray}
This point corresponds to the regime of ``pure'' advection by overall shaking of the container.

FP3: a fixed point with the following coordinates:
\begin{eqnarray}
    g^{*} = 0; \quad w^{*} = 0; \quad u^{*} = 8y/3; \quad x_{1}^{*}=x_{2}^{*} = \frac{1}{2}(\sqrt{13} - 1);\nonumber\\
    \lambda_i=\left\{y - \varepsilon, y - \varepsilon,y, \frac{47 + \sqrt{13}}{162} y, \frac{13 - \sqrt{13}}{18} y \right\}.
    \label{FP3}
\end{eqnarray}

This point corresponds to the regime of ``pure'' advection by the turbulent fluid described by the power-like term in (\ref{forceD}), (\ref{forceD1}); it is also the only one attractive for the physical values of parameters $\varepsilon = 2$, $y = 4$.\footnote{The eigenvalues $(y - \varepsilon)$ and $(13 - \sqrt{13})y/{18}$ in (\ref{FP3}) match the ones reported in \cite{RedBook} for passive advection; see expression (3.22) there. To see the similarity, note that $\sqrt{13}$ in (\ref{FP3}) is, in fact, $\sqrt{1+8(d+2)/d}$ taken at $d=4$.
For $d=2$ this gives $\sqrt{17}$, the result estimated as ``intriguing'' by~\cite{FNS}. Note that there is a misprint in eq.~(3.69) in the corresponding formula in \cite{FNS}: the correct right hand side is $(-1+\sqrt{17})/2$.}

Finally, marginal values of dimensionless ratios turned charges must be considered in order to check for any ``hidden'' fixed points. To do that, one must pass to a new set of charges that allow for $x_1$ and $x_2$ to approach zero or infinity and then to study the resulting system of $\beta$ functions. After performing that calculation, however, we found no new stable points. In particular, a point corresponding to the ``pure'' HK model with irrelevant environment motion was established in one of those systems but it had a pair of eigenvalues differing only in their sign which makes the point always unstable.

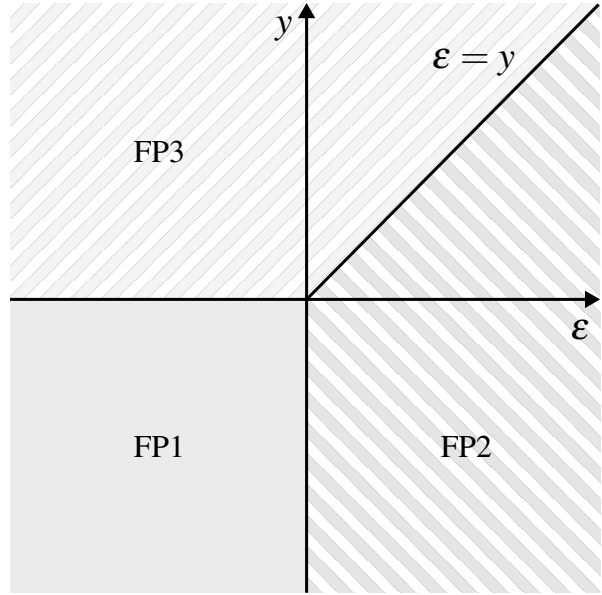
\begin{figure}[h]
    \centering
        \begin{overpic}[width=0.5\textwidth]{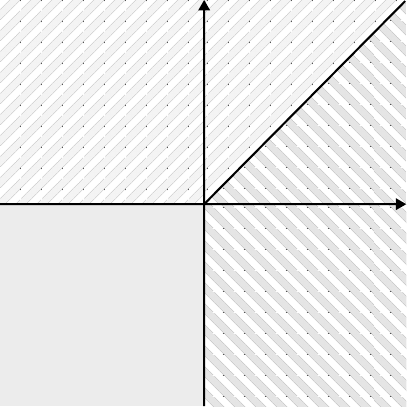}%
        \put(45,95){\large $y$}%
        \put(71.5,89){\large $\varepsilon=y$}%
        \put(21,23){FP1}
        \put(73,23){FP2}
        \put(21,73){FP3}
        \put(95,43){\large $\varepsilon$}
        \end{overpic}
    \caption{Stability regions of fixed points in the model (\ref{action_ve}) -- (\ref{action_ve2}).}
    \label{fig:enter-label}
\end{figure}

Regions of IR attractiveness of the fixed points FP1--FP3 in the $\varepsilon$--$y$ plane are shown on Figure~\ref{fig:enter-label}. They cover the entire plane with no gaps and overlaps, with straight boundaries between them. Although this result is derived within the one-loop approximation, it can be argued that it is not affected by the higher-order corrections and is exact. It is important here that the special cases $D_1=0$ and $D_2=0$ in the correlation function (\ref{forceD1}), (\ref{loc-nonloc}) give rise to two independent models that are multiplicatively renormalizable on their own; see Section~\ref{sec:level3}.

We can see that the most interesting case $\varepsilon=2$ ($d=2$) and $y\to 4$ (turbulent motion of the environment) corresponds to the fixed point 
FP3, where the HK nonlinearity and the local term in (\ref{loc-nonloc})
are IR irrelevant and one is left with the pure turbulent advection. 
This result may seem rather disappointing.\footnote{To prevent possible misunderstanding, it should be stressed that the bare couplings
$g$ and $w$ are kept fixed. It is the corresponding {\it invariant or running couplings}; see, {\it e.g.} Sections~1.29--1.32 in~\cite{Book3}, that asymptotically approach the fixed point with vanishing coordinates $w_*=g_*=0$. Thus, the HK interaction and the stirring survive in corrections to the leading-order terms of the IR asymptotic behaviour and, what is more important, contribute to determination of the stability regions.}
However, in the Section~\ref{idea}, we will discuss possible effects of the two-loop corrections and argue that they may result in appearance of additional, fully nontrivial fixed point, where both the HK nonlinearity and turbulent mixing are simultaneously relevant.

\section{Scaling behaviour and critical dimensions} \label{sec:CD}

By combining the RG equation at a fixed point with two equations reflecting the canonical scale invariance with respect to momentum and frequency, one arrives at the following expression for the  
critical dimensions $\Delta_{F}$ for a set of fields $F$
(see, {\it e.g.}~Section~2.1 in~\cite{RedBook}, Section~3.1 in~\cite{UFN} and Sections~5.16 and 6.7 in~\cite{Book3}):
\begin{eqnarray}
\Delta_{F} = d^k_{F} + \Delta_\omega d^{\omega}_{F} + \gamma_{F}^*.
\label{critical dimension}
\end{eqnarray}
Here $d^{k,\omega}_F$ are canonical dimensions of the fields $F$, $\gamma_{F}^*$ is the value of the anomalous dimension at the fixed point, and $\Delta_\omega$ is the critical dimension of frequency. 
For the model (\ref{action_ve}) -- (\ref{action_ve2}), the latter is expressed as
\begin{eqnarray}
\Delta_\omega = 2 - \gamma_{\nu}^*=2-\gamma_{3}^*,
\label{Domega}
\end{eqnarray}
while the critical dimensions of the fields $h,h',{\bf v},{\bf v'}$ are as follows:
\begin{eqnarray}
\Delta_h = 1 -\gamma_{\nu}^* + \gamma^* _h 
= 1 -\gamma_{3}^* + \gamma_{4}^*, \\
\Delta_{h'} = d- \Delta_{h}, \nonumber \\
\Delta_{\bf v} =  1 -\gamma_{\nu}^* + \gamma^* _{\bf v} 
= 1 -\gamma_{3}^*, \nonumber \\
 \Delta_{\bf v'} = d- \Delta_{\bf v}. \nonumber 
\end{eqnarray}
Here we used canonical dimensions from Table \ref{canonical dimensions} and relations between the anomalous dimensions (\ref{gamma}).

For the point FP1, the critical dimensions coincide with their canonical values since the anomalous dimensions $\gamma_{3}^*$ and $\gamma_{4}^*$ are equal to zero exactly (in all orders of the perturbation theory):
\begin{eqnarray}
\Delta_h=\Delta_{{\bf v}}=1, \quad 
\Delta_{h'}=\Delta_{{\bf v'}}=d-1, \quad \Delta_{\omega} =2.
\label{DimFP1}       
\end{eqnarray}

For the curve of points FP2 (including its endpoint FP2a), the critical dimensions are also determined exactly:
\begin{eqnarray}
\Delta_h=1-\epsilon/3, \quad \Delta_{h'}=d-1+\epsilon/3,
\label{DimFP2}       
\end{eqnarray}
\begin{eqnarray}
\Delta_{{\bf v}}=1-\epsilon/3, \quad \Delta_{{\bf v'}}=d-1+\epsilon/3, \quad \Delta_{\omega} =2-\epsilon/3.
\label{DimFP2a}       
\end{eqnarray}
Indeed, the expressions (\ref{DimFP2a}) are exact because the anomalous dimension $\gamma_{3}^*$ is known exactly for those points and they 
coincide with the dimensions derived in \cite{FNS} for the stirred NS equation. The anomalous dimension $\gamma_{4}^{*}$ is also known exactly and vanishes at all fixed points on this curve (see \cite{HKNS1} and Section \ref{idea} below).

For the point FP3, the critical dimensions are:
\begin{eqnarray}
\Delta_h=1-y/3, \quad \Delta_{h'}=d-1+y/3; \\
    \Delta_{{\bf v}}=1-y/3, \quad \Delta_{{\bf v'}}=d-1+y/3, \quad \Delta_{\omega} =2-y/3. 
\label{DimFP3} 
\end{eqnarray}
These expressions are exact to all orders in $y$ because $g^*=0$ and transversality of the velocity field ${\bf v}$ leads to the exact expression $\gamma_{4}^*=0$.

\section{Beyond the one-loop order: existence of fully nontrivial fixed point?}
\label{idea}

We calculated the renormalization constants (\ref{Z1})--(\ref{ZZZZ}) and the coordinates of the fixed points (\ref{FP1})--(\ref{FP3}) in the one-loop approximation. However, we managed to derive exact results for critical dimensions (\ref{DimFP1})--(\ref{DimFP3}). 
Thus, one can ask whether any new information can be revealed from the higher-order contributions? To answer this question, consider the anomalous dimension $\gamma_4$
in the two-loop approximation.

From the expressions for the $\beta$ functions (\ref{bx1}) and (\ref{bx2}), it follows that the equalities $\gamma_{1}^{*} = \gamma_{2}^{*} = \gamma_{3}^{*}$ should hold at any fixed point with nonvanishing coordinates $x_{1}^{*}$, $x_{2}^{*}$. If we are looking for a fixed point at which the HK nonlinearity is relevant ($g^{*} \neq 0$), the condition that $\beta_{g} = 0$ reduces to:
\begin{equation}
    \varepsilon - 3\gamma_{3}^{*} - 2\gamma_{4}^{*} = 0.
    \label{fpr}
\end{equation}
In a general situation, the system of equations $\beta_w=0$, $\beta_u=0$  has two solutions (see expressions (\ref{bw}), (\ref{bu})):
\begin{equation}
    u^{*} = 0,\quad \gamma_{3}^{*} = \varepsilon/3
    \label{sol1}
\end{equation}
and 
\begin{equation}
    w^{*} = 0,\quad \gamma_{3}^{*} = y/3.
    \label{sol2}
\end{equation}
The coordinates (\ref{sol1}) being substituted into equation (\ref{fpr}) give the obvious result
\begin{equation}
    \gamma_{4}^{*} = 0,
\end{equation}
discussed in details in \cite{HKNS1}. 

On the contrary, substituting the coordinates (\ref{sol2}) into (\ref{fpr}) results in the following equation for the coordinate $g^{*}\ne 0$:
\begin{equation}
    \gamma_{4}^{*} = (y-\varepsilon)/{2}.
\end{equation}
If $\gamma_{4}$ contains a nontrivial contribution at the two-loop level, it has to be of the form
\begin{equation}
    \gamma_{4} = A\, gw + B\, gu + \text{higher orders in couplings},
\end{equation}
where $A$ and $B$ are certain constant coefficients. Indeed, the terms of the form $g^{2}$ can arise only from the diagrams corresponding to the pure HK model, and therefore cancel each other due to the Galilean symmetry. On the other hand, the diagrams that could give rise to the $w^{2}$, $u^2$ and $uw$ terms, vanish due to transversality of the propagator $\langle vv \rangle_0$. Thus, for the fixed point with a nonvanishing coordinate 
$u^{*} = u_{1}^{*}\,y + u_{2}^{*}\,y^{2}+\dots$, obtained from the expressions (\ref{sol2}), one arrives at the two-loop solution 
\begin{equation}
    g^{*} = ({y - \varepsilon})/{2 B u^{*}},
    \label{hypo}
\end{equation}
that corresponds to a completely nontrivial fixed point at which both the HK nonlinearity and the turbulent advection are simultaneously relevant.

\section{Conclusion}\label{sec:Conc}

We studied a strongly anisotropic system described by the Hwa--Kardar continuous model of self-organized criticality (\ref{HKadjusted-fin}), 
(\ref{nabla-fin}), (\ref{forceC}) interacting with its moving fluid environment that is chosen to be isotropic and random, modeled by the stochastic NS equation (\ref{NS}), (\ref{forceD}). The latter involves a random stirring force of a general form (\ref{forceD1}), (\ref{loc-nonloc}) that includes both the uniform stirring of the system (the $D_1$ term) and the random motion that describes, in the limiting case $y\to4$, a turbulent fluid (the $D_2$ term).

The two coupled stochastic equations are rewritten as a field theoretic model (\ref{action_ve}), (\ref{action_ve2}) that we show to be multiplicatively renormalizable (\ref{act_r}), (\ref{renorm_param})  and logarithmic in the space dimension $d=4$ (see Table~\ref{canonical dimensions}). 

By calculating renormalization constants in the one-loop approximation (\ref{Z2})--(\ref{fff}), we arrived at the system of $\beta$ functions (\ref{b1})--(\ref{b4}) that allowed us to find the attractors of the RG equations (\ref{FP1})--(\ref{FP3}) in the five-dimensional space of the model parameters. They correspond to the following regimes of infrared/long-range asymptotic behaviour of the system (or rather of the correlation and response functions describing it): ordinary diffusion (a two-dimensional surface of Gaussian fixed points (\ref{FP1})); advection by the turbulent fluid (the single fixed point (\ref{FP3}), the Hwa--Kardar self-interaction is irrelevant there) and a crossover regime (one-dimensional curve of fixed points (\ref{FP2}), (\ref{FP2a})) where both the Hwa--Kardar nonlinearity and uniform stirring are simultaneously relevant.  

The exponent $y$ and the dimension of space $d$ enter the eigenvalues $\lambda$ of the stability matrix $\Omega$ and therefore determine the regions of attractiveness of the corresponding IR asymptotic regimes; the most realistic cases $d=2$ or $d=3$ and $y\to 4$ correspond to the single fixed point (\ref{FP3}). The critical dimensions of the frequency and those of the basic fields are found exactly (\ref{DimFP1})--(\ref{DimFP3}).

However, analysis of possible two-loop corrections suggests the existence of another nontrivial fixed point (\ref{hypo}), not detectable in the one-loop expressions (\ref{b1})--(\ref{b4}) for the $\beta$ functions. That point corresponds to the regime where the HK nonlinearity and advection by the turbulent fluid are simultaneously relevant. Careful analysis of this fixed point, its stability and critical dimensions requires rather cumbersome two-loop calculation, which lies beyond the scope of the present study.

This result is particularly interesting compared to previous findings \cite{AK19,Serov1,Serov2,WeU,WeU2,HKNS1,HKNS2,HKNS3}; firstly, the pattern of stability obtained from the one-loop calculations is in line with that found in the studies of the Kardar--Parizi--Zhang (KPZ) equation coupled to the NS equation \cite{KPZ-NS-JPA,Scripta}. Indeed, when HK or KPZ models are put in the environment described by the NS equation with the $D_1$ term, the regime of the pure nonlinearity becomes unstable while the crossover regime corresponds to the realistic values of the parameters \cite{HKNS1,HKNS2,KPZ-NS-JPA}. However, if the NS equation is extended to include the $D_2$ term, the pure regime corresponding to that term\footnote{With a caveat that in the case of KPZ, the pure regime includes the $D_1$ term too due to the model not being renormalizable otherwise.} ``wins'' \cite{Scripta}.

This is similar to how the pure regime of the Kraichnan's velocity ensembles wins against the HK nonlinearity in works \cite{WeU,WeU2}. That is to say, realistic values of the parameters correspond to the regime of pure isotropic turbulent advection.

On the contrary, if we use the anisotropic Avellaneda--Majda velocity ensemble and its generalizations, the realistic values correspond to the special crossover regimes that are restricted by a relation between $\varepsilon$ and the ensemble exponent \cite{AK19,Serov1,Serov2}. Note that in the present model, there is also a crossover (between the HK nonlinearity and the $D_1$ term) for the case $\varepsilon=y$. 
 
From a physical point of view, it would seem plausible for isotropic velocity fields describing turbulence to ``wash away'' anisotropic nonlinearity of the HK equation while anisotropic velocity fields or random motion of the environment play on a more of a level field with it. However, if our arguments about two-loop fixed point hold, that new crossover regime might contradict the pattern above. But this point stability region is just as likely not to correspond to the most realistic values of the parameters. So for now, we cannot solve the uncertainty.

Another possible direction of research is an extension of the model (\ref{HKadjusted-fin}). In the above study, we followed the original logic of Hwa and Kardar by keeping the premise of continuity equation and introducing terms into the current (\ref{current_long}) not higher than the second order in the fields: this can be justified {\it a posteriori} by canonical dimensions analysis. The requirements of incompressibility and the Galilean symmetry (\ref{galilean}) impose severe restrictions on (\ref{current_long}), such that the only term $\partial ({\bf v} h) = ({\bf v}\partial)h$  survives. It is naturally interpreted as a part of the Lagrangian (Galilean covariant) derivative (\ref{nabla-fin}) for a conserved density field $h$. If (\ref{galilean}) is relaxed to transformations along the preferred direction, ${\bf u}={\bf n}u$, the additional anisotropic advection term $({\bf n}\partial)({\bf n}{\bf v}h)$ becomes possible. If the Galilean symmetry is abandoned at all, the two last $\beta$ terms in (\ref{current_long}) should also be included. Clearly, this may strongly affect the IR behaviour of the extended model. In particular, an interesting issue is the possible restoration of isotropy and Galilean invariance for certain IR scaling regimes; in this connection see, e.g.~\cite{restobar}. This work is in progress.

\section*{Acknowledgments}

This research was funded by the Russian Science Foundation grant number 
24-22-00220, https://rscf.ru/en/project/24-22-00220/.

\section*{Appendix: One-loop calculations \label{sect5}} 

In this Appendix, we will briefly explain technique used to calculate renormalization constants and give a comment on triviality of the constant $Z_{4}$ in (\ref{ZZZZ}).

The building blocks of the perturbation theory for the model~(\ref{action_ve}), (\ref{action_ve2}) comprise four bare propagators and three bare vertices.
The bare propagators in the frequency-momentum $(\omega-{\bf k})$ representation are:
\begin{eqnarray}
    \langle hh' \rangle_0 &= \langle h'h \rangle_0^{*} = \frac{1}{-{\rm i}\omega + \stdepsilon({\bf k})}, \quad \
    \langle h'h' \rangle_0= 0, \quad \
    \langle hh \rangle_0 = \frac{C}{\omega^2+\stdepsilon^2({\bf k})}, \nonumber \\
    \langle v_i v'_j \rangle_0 &= \langle  v'_i v_j \rangle_0^{*} = \frac{P_{ij}({\bf k})}{-{\rm i}\omega + \nu k^2}, \quad
    \langle v'_i v'_j \rangle_0 = 0, \,\,\,\,\,\,
    \langle  v_i v_j \rangle_0 = \frac{D(k)\, P_{ij}({\bf k})}{\omega^2 + {\nu}^2 k^4},
    \label{propagators}
\end{eqnarray}
where $\stdepsilon({\bf k}) = \nu_{\parallel}{ k}_{\parallel}^2 + \nu_{\perp}{\bf k}_{\perp}^2$. The propagators $\langle h'h' \rangle_{0}$ and $\langle v'_i v'_j \rangle_{0}$ are equal to zero due to causality considerations (see, {\it e.g.} Sections~5.3, 5.4 in~\cite{Book3}). The same applies to exact dressed propagators, so no diagrams will give a contribution to the corresponding Green functions.
We will use the following graphical notations for the propagators when drawing Feynman diagrams:
\begin{eqnarray}
     \langle hh' \rangle_{0} = \hhstroke, & \quad \langle hh \rangle_{0} = \lineonly, \quad
     \nonumber
     \\
     \langle v_i v'_j \rangle_{0} = \vvstrokeHKNS, & \quad \langle v_i v_j \rangle_{0} = \lineonlyboson.
     \nonumber
     \label{vertices}
\end{eqnarray}
The three interactions terms in (\ref{action_ve}), (\ref{action_ve2})
\begin{eqnarray}
-\frac{1}{2}h'\partial_\parallel h^2=\frac{1}{2} J h'hh, 
\quad
-h'({\bf v} {\bf  \partial})h= {\tilde J}_s h' v_s  h, 
\quad
-{\bf v'}({\bf v}{\bf  \partial}){\bf v}= \frac{1}{2} I_{j,sn}  v'_j v_s v_n
\end{eqnarray}
correspond to the three possible vertices with the vertex factors
\begin{eqnarray}
    \hhhstrokeVertex = J = {\rm i}k_\parallel, \quad \hvhstrokeVertex = {\tilde J}_s = {\rm i} k_s, \nonumber\\ \vvvstrokeVertex = I_{j,sn}= {\rm i}(k_n\, \delta_{sj} + k_s\, \delta_{nj}).
\end{eqnarray}
Then the one-loop approximation for the 1-p irreducible Green's functions containing UV divergences to be eliminated is:
\begin{eqnarray}
     \langle h h' \rangle_{\rm 1-ir} &=& {\rm i}\omega - \nu_{\parallel} p_{\parallel}^2 Z_1 - \nu_{\bot} p_{\bot}^2 Z_2 + \nonumber\\ & + &\hhstrokeFirst \ + \hhstrokeSecond \ \ , 
    \label{hh'}
\end{eqnarray}
\vspace{-1.0ex}
\begin{eqnarray}
     \langle  { {\bf v}} { {\bf v}}' \rangle_{\rm 1-ir} = {\rm i}\omega - \nu p^2 Z_3 \ + \vvstroke \ \ , 
     \label{vv'}
\end{eqnarray}

\begin{eqnarray}
\langle h'hh \rangle_{1-ir} = 
{\rm i}p_{\parallel}Z_{4} + \hhhstrokeone + \hhhstroketwo + \hhhstrokethree \nonumber \\ 
+\hhhstrokefour + \hhhstrokefive + \hhhstrokesix\raisebox{-2ex}[1.5cm][0cm]{.}
\label{h'hhir}
\end{eqnarray}
\\
In order to extract and eliminate UV divergences we will use dimensional and analytical regularization along with the minimal subtraction (MS) scheme. Since the dimensional regularization does not distinguish between UV and IR divergences, an additional IR regularization of the diagrams should be provided. On the other hand, within the MS scheme, the explicit form of renormalization constants does not depend on precise type of IR regularization (because it is capable of changing only the finite parts of the momentum integrals), so that in each particular diagram we are free to choose IR regularization most suitable from a technical standpoint.

For the diagrams in (\ref{hh'}) and (\ref{h'hhir}), the most convenient choice is sharp IR-cutoff. The cut-off parameter has the meaning of the inverse scale of the largest structures in system \cite{DM79,UFN,RedBook,Book3}, its precise value is not important for our analysis. For the diagram in (\ref{vv'}), because of computational reasons, it is more convenient to provide IR regularization not by imposing a cutoff but by holding the non-vanishing external momentum throughout the calculation.

Since there are no frequency divergences in the model, all diagrams in (\ref{hh'})--(\ref{h'hhir}) can be evaluated at vanishing external frequencies, while the integral over loop frequency can be calculated by residues.

In order to calculate divergent parts of the resulting momentum integrals it is sufficient to employ three well-know formulae, namely the generalized Feynman formula:
\begin{equation}
 A_1^{-\lambda_1}\dots A_n^{-\lambda_n}=\frac{\Gamma(\sum_{i=1}^{n}\lambda_i)}{\prod_{i=1}^{n}\Gamma(\lambda_i)}\int_0^1\dots\int_0^1 du_1\dots du_n \frac{\delta\left(\sum_{i=1}^n u_i - 1\right)\prod_{i=1}^nu_i^{\lambda_i-1}}{\left[\sum_{i=1}^n A_i u_i\right]^{\sum_{i=1}^n\lambda_i}}, 
\label{Feynman}
\end{equation}
the Schwinger representation:
\begin{eqnarray}
   A^{-\lambda} = \frac{1}{\Gamma \left(\lambda \right)} \int_{0}^{\infty} dt \, t^{\lambda-1} \exp\{-tA\}, 
    \label{Schwinger}
\end{eqnarray}
and the $d$-dimensional Gaussian integral:
\begin{equation}
    \int d^{d}{\bf k} \exp{\left\{-\frac{1}{2}{\bf k} M {\bf k} + {\bf q}{\bf k}\right\}} = \det\left\{\frac{M}{2\pi}\right\}^{-1/2} \exp{\left\{\frac{1}{2}{\bf q}M^{-1}{\bf q}\right\}}; \quad 
\label{Gauss}
\end{equation}
see, {\it e.g.} Section~3.15 in ~\cite{Book3}.

Let us first discuss diagrams contributing to $\langle h'hh \rangle_{1-ir}$. All the diagrams in (\ref{h'hhir}) can be divided in two groups: those without velocity field propagators (first row) and those containing velocity field propagators (second row). Each diagram in the first group contain nontrivial UV divergences. However, due to the Galilean symmetry of the original HK model, they  cancel each other and invest no contribution to $Z_{4}$. The diagrams of the second group are rendered finite by the transversality of propagator $\langle v_{i}v_{j} \rangle_{0}$. The reasoning is as follows. After factoring out the longitudinal part of the external momentum, the remaining integral diverges logarithmically. Within the MS scheme, logarithmic UV divergences are independent of external momenta, allowing one to set latter to zero. The numerator of each second-group diagram then necessarily contains a contraction of the form $P_{ij}(k)k_{i} = 0$, where $P_{ij}$ comes from the velocity propagator, and $k_{i}$, being loop momenta, originates from the second vertex of (\ref{vertices}). Hence, within the one-loop approximation the constant $Z_{4} = 1$ receives no contribution and remains trivial.

The details of the calculation of the only diagram contributing to (\ref{vv'}) can be found in the Appendix of \cite{KPZ-NS-JPA} (calculation of the $A_{9}$ diagram). The calculation algorithm is rather straightforward. First the frequency integral is computed and contractions in numerator of the integrand are performed. Then the denominator of the integrand is expanded in Taylor series in powers of external momenta and UV finite terms of the expansion are discarded. By computing the remaining integral of the pure power of the loop momentum modulus, the final result is obtained:
\begin{equation}
    \vvstroke = 
     - \frac{1}{8\pi^2} \nu p^2\frac{1}{ 8}  \left(\frac{w}{\epsilon} + \frac{u}{y}\right) + \text{finite part}.
\end{equation}
By redefining charges $w/8\pi^2\rightarrow w$ and $u/8\pi^2\rightarrow u$, we thereby arrive at the expression (\ref{Z3}) for the renormalization constant $Z_{3}$.

The UV divergent part of the first diagram in (\ref{hh'}) can be calculated by repeating the same steps (for the details see calculation of the diagram $D_{2}$ in the Appendix of \cite{Serov2}). The only difference is that we have to perform additional change of variables of the form $k_{\parallel} \to \nu_{\parallel}^{-1/2}k_{\parallel}$ and $k_{\bot} \to \nu_{\bot}^{-1/2}k_{\bot}$ in the momentum integral, which allows to reduce corresponding integral to the integral of $O(d)$-symmetric integrand. The final expression:
\begin{equation}
      \hhstrokeFirst = -\frac{1 }{8\pi^2} \nu_{\parallel}  p_\parallel^2 \frac{3}{16}\frac{ g}{\epsilon} + \text{finite part}
\label{hktwopoint}
\end{equation}
with redefinition $g/8\pi^2\rightarrow g$ contributes to the renormalization constant $Z_{1}$, producing first term in the brackets in (\ref{Z1}).

The second diagram in (\ref{hh'}) is the most challenging computationally, therefore, we will describe computation of its UV-divergent part in detail:
\begin{eqnarray}
    \hhstrokeSeconddetail\,
    = \int \frac{d\omega\,d^d {\bf k}}{(2\pi)^{d+1}}\frac{ \left(D_{1} + D_{2} \, k^{4-d-y}\right) P_{ab} ({\bf k}) \, 
    i (p-k)_a i p_b
    }{(\omega^2+\nu ^2 k^4)(-i\omega+\stdepsilon({\bf p}-{\bf k}))} = \nonumber \\
    =-\frac{1}{2}\int \frac{d^d {\bf k}}{(2\pi)^d} \frac{ \left(D_{1} + D_{2} \, k^{4-d-y}\right) P_{ab} ({\bf k}) \, 
    p_a p_b
    }{\nu k^2(\nu k^{2}+\stdepsilon({\bf p}-{\bf k}))} = \widetilde{D}_1 + \widetilde{D}_2.
\label{hh' light 1}
\end{eqnarray}
It is convenient to split this diagram into a sum of two contributions: $\widetilde{D}_1$, the part proportional to $D_{1}$, and $\widetilde{D}_2$, the part proportional to $D_{2}$. Calculation of the divergent part of the integral $\widetilde{D}_1$ has been described in detail in \cite{HKNS1}. Here we will compute it using a different method, consistent with how we will calculate the divergent part of $\widetilde{D}_2$. The result of \cite{HKNS1} herewith will be reproduced.

In explicit notation:
\begin{eqnarray}
        \widetilde{D}_{1} &=& - \frac{D_{1}}{2 \nu_0} \int \frac{d^{d} {\bf k}}{(2\pi)^d} \frac{ P_{ab} ({\bf k}) p_a p_b }{k^2 (\nu k^{2} + \stdepsilon({\bf p}-{\bf k}))}.
\label{D1wintegrated}
\end{eqnarray}
After performing contraction in the numerator $P_{ab} ({\bf k}) p_a p_b = p^{2} - k^{-2}(p_{\parallel}k_{\parallel} + {\bf p}_{\bot}{\bf k}_{\bot})^{2}$, it is convenient to once again split (\ref{D1wintegrated}) into the sum of two separate integrals: one containing isotropic term and the other containing anisotropic term in the numerator:
\begin{eqnarray}
        \widetilde{D}_{1} &=&- \frac{D_{1}p^{2}}{2 (2\pi)^d} I_{1} +  \frac{D_{1}\nu}{2 (2\pi)^d} I_{2},
\label{D1split}
\end{eqnarray}
where
\begin{eqnarray}
I_{1} &=& \int d^{d} {\bf k} \frac{ 1 }{\nu k^2 (\nu k^{2} + \stdepsilon({\bf p}-{\bf k}))},
\label{I1}        
\end{eqnarray}
\begin{eqnarray}
I_{2} &=& \int d^{d} {\bf k} \frac{ (p_{\parallel}k_{\parallel} + {\bf p}_{\bot}{\bf k}_{\bot})^{2} }{\nu^{2}k^4 (\nu k^{2} + \stdepsilon({\bf p}-{\bf k}))}.
\label{I2}
\end{eqnarray}

Let us begin by computing the simpler integral (\ref{I1}). Employing the Feynman formula (\ref{Feynman}), we obtain:
\begin{eqnarray}
    I_{1}=\int_{0}^{1}dx \int d^{d} {\bf k} \frac{ 1 }{(\nu k^2x + (1-x)(\nu k^{2} + \stdepsilon({\bf p}-{\bf k})))^{2}}.
\end{eqnarray}
Making use of the Schwinger parametrization (\ref{Schwinger}), the last expression can be rewritten as
\begin{eqnarray}
    I_{1}=\int_{0}^{1}dx \int_{0}^{\infty} dt \int d^{d} {\bf k}\ t \exp{[-t(\nu k^2 + (1-x)\stdepsilon({\bf p}-{\bf k}))]}.
\end{eqnarray}
The momentum integral is now of Gaussian type and can be straightforwardly computed by applying formula (\ref{Gauss}):
\begin{eqnarray}
    I_{1}= \pi^{d/2}\int_{0}^{1}dx \int_{0}^{\infty} dt \ t^{1-d/2}\frac{(\nu+(1-x)\nu_{\bot})^{(1 - d)/2}}{(\nu+(1-x)\nu_{\parallel})^{1/2}}\exp{[-tf({\bf p})]},
\end{eqnarray}
where we have introduced shorthand notation:
\begin{eqnarray}
    f({\bf p}) = (1-x)\left\{(\nu_{\parallel}p_{\parallel}^{2}+\nu_{\bot}{\bf p}_{\bot}^{2}) - \frac{\nu_{\parallel}^{2}p_{\parallel}^{2}}{\nu+(1-x)\nu_{\parallel}}- \frac{\nu_{\bot}^{2}p_{\bot}^{2}}{\nu+(1-x)\nu_{\bot}}\right\}.
\end{eqnarray}
The integral over the parameter $t$ is calculated by reapplying (\ref{Schwinger}):
\begin{eqnarray}
    I_{1} = \pi^{1/2}\,\Gamma(2-{d}/{2})\int_{0}^{1}dx \frac{(\nu+(1-x)\nu_{\bot})^{\frac{1 - d}{2}}}{(\nu+(1-x)\nu_{\parallel})^{\frac{1}{2}}} f^{\frac{4-d}{2}}({\bf p}).
\end{eqnarray}
The function $f({\bf p})$ serves as an IR regulator, ensuring that the final expression remains free of IR divergences. Upon performing a Laurent expansion of the last expression in $\varepsilon$, the UV divergence manifests as a simple pole in the leading singular term:
\begin{eqnarray}
    I_{1} = \frac{2\pi^{2}}{\varepsilon}\int_{0}^{1}dx (\nu+(1-x)\nu_{\bot})^{-3/2}(\nu+(1-x)\nu_{\parallel})^{-1/2} + \mathcal{O}(\varepsilon^{0}).
\end{eqnarray}
Evaluating the integral over $x$, we arrive at explicit expression for UV divergent part of $I_{1}$:
\begin{eqnarray}
    I_{1} = \frac{4\pi^{2}}{\varepsilon}\frac{\sqrt{(\nu+\nu_{\parallel})/(\nu+\nu_{\bot})}-1}{\nu(\nu_{\parallel}-\nu_{\bot})} + \text{finite part}.
    \label{I1final}
\end{eqnarray}

The calculation of the UV divergent part of $I_{2}$ proceeds along the same lines. As a first step we employ the Feynman formula:
\begin{eqnarray}
    I_{2}=\int_{0}^{1}dx \int d^{d} {\bf k} \frac{ (p_{\parallel}k_{\parallel} + {\bf p}_{\bot}{\bf k}_{\bot})^{2} }{(\nu k^2x + (1-x)(\nu k^{2} + \stdepsilon({\bf p}-{\bf k})))^{3}},
\end{eqnarray}
and the Schwinger representation:
\begin{eqnarray}
    I_{2}=\int_{0}^{1}dx \int_{0}^{\infty} dt \int d^{d} {\bf k}\ t^{2} (p_{\parallel}k_{\parallel} + {\bf p}_{\bot}{\bf k}_{\bot})^{2}\exp{[-t\,\Xi({\bf p}, {\bf k})]}.
\end{eqnarray}
In the last expression we have introduced the following abbreviated notation:
\begin{eqnarray}
    &\Xi({\bf p}, {\bf k}) = \nu k^2 + (1-x)\stdepsilon({\bf p}-{\bf k})) = \nu k^2\nonumber \\
    &+ (1-x)\Big((\nu_{\parallel}k^{2}_{\parallel}+\nu_{\bot}{\bf k}^{2}_{\bot}) - 2(\nu_{\parallel}p_{\parallel}q_{\parallel}+\nu_{\bot}{\bf p}_{\bot}{\bf k}_{\bot}) + (\nu_{\parallel}p^{2}_{\parallel}+\nu_{\bot}{\bf p}^{2}_{\bot})\Big).
\end{eqnarray}
It is also convenient to define:
\begin{eqnarray}
    &\Xi_{\alpha, \beta}({\bf p}, {\bf k}) = \nu k^2\nonumber + (1-x)\Big((\nu_{\parallel}k^{2}_{\parallel}+\nu_{\bot}{\bf k}^{2}_{\bot}) +(\nu_{\parallel}p^{2}_{\parallel}+\nu_{\bot}{\bf p}^{2}_{\bot})\Big) + \nonumber \\
    &+\alpha p_{\parallel}q_{\parallel} + \beta{\bf p}_{\bot}{\bf k}_{\bot}, 
\end{eqnarray}
so that
\begin{eqnarray}
    &\Xi_{\alpha, \beta}({\bf p}, {\bf k})\Bigg|_{\substack{\alpha = -2(1-x)\nu_{\parallel} \\ \beta = -2(1-x)\nu_{\bot}}} = \Xi_{\alpha, \beta}({\bf p}, {\bf k})\Bigg|_{*} = \Xi({\bf p}, {\bf k}).
\end{eqnarray}
Then by introducing differential operator $L_{\alpha, \beta} = \partial_{\alpha}^{2} + \partial_{\beta}^{2} + 2\partial_{\alpha}\partial_{\beta}$, we can rewrite $I_{2}$ as
\begin{eqnarray}
    I_{2}=\int_{0}^{1}dx \int_{0}^{\infty} dt \int d^{d} {\bf k}\  L_{\alpha, \beta}\exp{[-t(\Xi_{\alpha, \beta}({\bf p}, {\bf k}))]}\Bigg|_{*}.
\end{eqnarray}
In the last expression, the integral over momentum becomes Gaussian and can be calculated using (\ref{Gauss}):
\begin{eqnarray}
    I_{2}=\int_{0}^{1}dx \int_{0}^{\infty} dt \ \Big(\frac{\pi}{t}\Big)^{d/2}\frac{(\nu+(1-x)\nu_{\bot})^{(1 - d)/2}}{(\nu+(1-x)\nu_{\parallel})^{1/2}}L_{\alpha, \beta}\exp{[-tf_{\alpha, \beta}({\bf p})]}\Bigg|_{*},
\end{eqnarray}
where we have introduced another shorthand notation:
\begin{eqnarray}
    f_{\alpha,\beta}({\bf p}) = (1-x)(\nu_{\parallel}p_{\parallel}^{2}+\nu_{\bot}{\bf p}_{\bot}^{2}) + \frac{\alpha^{2}p_{\parallel}^{2}}{4(\nu+(1-x)\nu_{\parallel})}+ \frac{\beta^{2}p_{\bot}^{2}}{4(\nu+(1-x)\nu_{\bot})}.
\end{eqnarray}
The operator $L_{\alpha, \beta}$ acts explicitly as follows:
\begin{eqnarray}
    L_{\alpha, \beta}\exp{[-tf_{\alpha, \beta}({\bf p})]} = \Big[\frac{p^{2}_{\parallel}t}{2(\nu+(1-x)\nu_{\parallel})} + \frac{p^{2}_{\bot}t}{2(\nu+(1-x)\nu_{\bot})} + \nonumber\\ 
    + \frac{\alpha^{2}p^{4}_{\parallel}t^{2}}{4(\nu+(1-x)\nu_{\parallel})^{2}} + \frac{\alpha^{2}p^{4}_{\bot}t^{2}}{4(\nu+(1-x)\nu_{\bot})^{2}} + \nonumber \\ 
    + \frac{\alpha \beta p^{2}_{\parallel}p^{2}_{\bot}t^{2}}{2(\nu+(1-x)\nu_{\parallel})(\nu+(1-x)\nu_{\parallel})}\Big]\exp{[-tf_{\alpha, \beta}({\bf p})]}.
\label{LonExp}
\end{eqnarray}
The last three terms are proportional to the fourth power of the external momenta; consequently, they contribute exclusively to the UV-finite part of $I_{2}$ and can thus be safely discarded at this stage. Taking the integral over $t$ of what is remaining and setting $\alpha = -2(1-x)\nu_{\parallel}$, $\beta = -2(1-x)\nu_{\bot}$, we obtain:
\begin{eqnarray}
    I_{2}=\frac{\pi^{d/2}\Gamma(2-d/2)}{2}\int_{0}^{1}dx \ \Bigg[\frac{p_{\parallel}^{2}x(\nu+(1-x)\nu_{\bot})^{(1 - d)/2}}{(\nu+(1-x)\nu_{\parallel})^{3/2}} + \nonumber \\
    +\frac{p_{\bot}^{2}x(\nu+(1-x)\nu_{\bot})^{(-1 - d)/2}}{(\nu+(1-x)\nu_{\parallel})^{1/2}}\Bigg]f({\bf p})^{2-d/2}+\text{finite part.}
\end{eqnarray}
Applying Laurent expansion with respect to $\varepsilon$ to the last expression yields the ultraviolet-divergent contribution of the form:
\begin{eqnarray}
    I_{2}=\frac{\pi^{2}}{2\varepsilon}\int_{0}^{1}dx\Bigg[\frac{p_{\parallel}^{2}x(\nu+(1-x)\nu_{\bot})^{\frac{-3}{2}}}{(\nu+(1-x)\nu_{\parallel})^{\frac{3}{2}}}     +\frac{p_{\bot}^{2}x(\nu+(1-x)\nu_{\bot})^{\frac{-5}{2}}}{(\nu+(1-x)\nu_{\parallel})^{\frac{1}{2}}}\Bigg]+\mathcal{O}(\varepsilon^{0})
     \label{I2expanded}
\end{eqnarray}
By evaluating integral over $x$, we arrive at:
\begin{eqnarray}
    I_{2}=\frac{\pi^{2}}{\varepsilon}\Bigg[\frac{p^{2}_{\parallel}}{\nu^{2}(2\nu + \nu_{\parallel} + \nu_{\bot} + 2\sqrt{(\nu+\nu_{\parallel})(\nu + \nu_{\bot})})} + \nonumber \\
    p^{2}_{\bot}\frac{\nu_{\bot} - 3\nu_{\parallel}-2\nu+2\left(\nu+\nu_{\parallel}\sqrt{(\nu+\nu_{\parallel})/
    (\nu+\nu_{\bot}})\right)}{3\nu^{2}(\nu_{\parallel}-\nu_{\bot})^{2}}\Bigg] + \text{finite part.}
    \label{I2final}
\end{eqnarray}

Now let us compute a somewhat more involved integral:
\begin{eqnarray}
        \widetilde{D}_{2} &=& - \frac{D_{2}}{2 \nu_0} \int \frac{d^{d} {\bf k}}{(2\pi)^d} \frac{ k^{4-d-y}P_{ab} ({\bf k}) p_a p_b }{k^2 (\nu k^{2} + \stdepsilon({\bf p}-{\bf k}))}.
\label{D2wintegrated}
\end{eqnarray}
In analogy with the previous case, we decompose the expression into terms containing isotropic and anisotropic numerator contributions:
\begin{eqnarray}
        \widetilde{D}_{2} &=&- \frac{D_{2}p^{2}\nu^{\frac{d+y-4}{2}}}{2 (2\pi)^d} J_{1} +  \frac{D_{2}\nu^{\frac{d+y-2}{2}}}{2 (2\pi)^d} J_{2},
\label{D2split}
\end{eqnarray}
 {with}
\begin{eqnarray}
J_{1} &=& \int d^{d} {\bf k} \frac{ 1 }{(\nu k^2)^{\frac{d+y-2}{2}} (\nu k^{2} + \stdepsilon({\bf p}-{\bf k}))},
\label{J1}        
\end{eqnarray}
\begin{eqnarray}
J_{2} &=& \int d^{d} {\bf k} \frac{ (p_{\parallel}k_{\parallel} + {\bf p}_{\bot}{\bf k}_{\bot})^{2} }{\nu^{2}k^4 ((\nu k^{2})^{\frac{d+y}{2}} + \stdepsilon({\bf p}-{\bf k}))}.
\label{J2}
\end{eqnarray}
The calculation follows the same methodology. We begin with $J_{1}$: applying the Feynman formula (\ref{Feynman}) yields
\begin{eqnarray}
    J_{1}=\frac{\Gamma(\frac{d+y+2}{2})}{\Gamma(\frac{d+y}{2})}\int_{0}^{1}dx \int d^{d} {\bf k} \frac{ x^{\frac{d+y-2}{2}} }{(\nu k^2x + (1-x)(\nu k^{2} + \stdepsilon({\bf p}-{\bf k})))^{\frac{d+y}{2}}}.
\end{eqnarray}
Using (\ref{Schwinger}), we recast this expression in the form
\begin{eqnarray}
    J_{1}=\frac{1}{\Gamma(\frac{d+y-2}{2})}\int_{0}^{1}dx \int_{0}^{\infty} dt \int d^{d} {\bf k}\ x^{\frac{d+y-4}{2}}t^{\frac{d+y-2}{2}} \exp{[-t\Xi({\bf p}, {\bf k})]}.
\end{eqnarray}
By evaluating Gaussian momentum integral
\begin{eqnarray}
    J_{1}=\frac{\pi^{d/2}}{\Gamma(\frac{d+y-2}{2})}\int_{0}^{1}dx \int_{0}^{\infty} dt \ x^{\frac{d+y-4}{2}}t^{\frac{y-2}{2}}\frac{(\nu+(1-x)\nu_{\bot})^{\frac{1 - d}{2}}}{(\nu+(1-x)\nu_{\parallel})^{1/2}}\exp{[-tf({\bf p})]}
\end{eqnarray}
and the integral over parameter $t$, we arrive at
\begin{eqnarray}
    J_{1} = \frac{\pi^{d/2}\Gamma(\frac{y}{2})}{\Gamma(\frac{d+y-2}{2})}\int_{0}^{1}dx\ x^{\frac{d+y-4}{2}}\frac{(\nu+(1-x)\nu_{\bot})^{\frac{1 - d}{2}}}{(\nu+(1-x)\nu_{\parallel})^{\frac{1}{2}}} f^{\frac{-y}{2}}({\bf p}).
\end{eqnarray}
The function $f(p)$ again serves as an IR regulator. However, in this case we perform a joint Laurent expansion in both the $y$ and $\varepsilon$ parameters. The negative power term exclusively captures the UV-divergent contribution to the expression
\begin{eqnarray}
    J_{1} = \frac{2\pi^{2}}{y}\int_{0}^{1}dx (\nu+(1-x)\nu_{\bot})^{-3/2}(\nu+(1-x)\nu_{\parallel})^{-1/2} + \mathcal{O}(\varepsilon^{0},y^{0}).
\end{eqnarray}
By evaluating the integral over $x$, we arrive at
\begin{eqnarray}
    J_{1} = \frac{4\pi^{2}}{y}\frac{\sqrt{\frac{\nu+\nu_{\parallel}}{\nu+\nu_{\bot}}}-1}{\nu(\nu_{\parallel}-\nu_{\bot})} + \text{finite part.}
    \label{J1final}
\end{eqnarray}
The Feynman parametrization for the last integral reads
\begin{eqnarray}
    J_{2}=\frac{\Gamma(\frac{d+y}{2})}{\Gamma(\frac{d+y-2}{2})}\int_{0}^{1}dx \int d^{d} {\bf k} \frac{ x^{\frac{d+y-2}{2}}(p_{\parallel}k_{\parallel} + {\bf p}_{\bot}{\bf k}_{\bot})^{2} }{(\nu k^2x + (1-x)(\nu k^{2} + \stdepsilon({\bf p}-{\bf k})))^{\frac{d+y+2}{2}}},
\end{eqnarray}
while the Schwinger representation is given by
\begin{eqnarray}
    J_{2}=\int_{0}^{1}dx \int_{0}^{\infty} dt \int d^{d} {\bf k}\ \frac{x^{\frac{d+y-2}{2}}t^{\frac{d+y}{2}}}{\Gamma(\frac{d+y-2}{2})} (p_{\parallel}k_{\parallel} + {\bf p}_{\bot}{\bf k}_{\bot})^{2}\exp{[-t(\Xi({\bf p}, {\bf k}))]},
\end{eqnarray}
with the same function $\Xi({\bf p}, {\bf k})$ as previously. Similar to the case of integral $I_{2}$, it is convenient to introduce the same function $\Xi_{\alpha, \beta}({\bf p}, {\bf k})$ and express the anisotropic momentum factor preceding the exponential as a differential operator $L_{\alpha, \beta}$, identical to the previous case
\begin{eqnarray}
    J_{2}=\int_{0}^{1}dx \int_{0}^{\infty} dt \int d^{d} {\bf k}\ \frac{x^{\frac{d+y-2}{2}}t^{\frac{d+y-4}{2}}}{\Gamma(\frac{d+y-2}{2})} L_{\alpha, \beta}\exp{[-t(\Xi_{\alpha, \beta}({\bf p}, {\bf k}))]}\Bigg|_{*}.
\end{eqnarray}
Taking Gaussian integral over momenta we get
\begin{eqnarray}
    J_{2}=\int_{0}^{1}dx \int_{0}^{\infty} dt \ \frac{\pi^{\frac{d}{2}}x^{\frac{d+y-2}{2}}t^{\frac{y-4}{2}}}{\Gamma(\frac{d+y-2}{2})}\frac{(\nu+(1-x)\nu_{\bot})^{\frac{1 - d}{2}}}{(\nu+(1-x)\nu_{\parallel})^{\frac{1}{2}}}L_{\alpha, \beta}\exp{[-tf_{\alpha, \beta}({\bf p})]}\Bigg|_{*}.
\end{eqnarray}
The operator $L_{\alpha, \beta}$ acts on the integration result exactly as before in (\ref{LonExp}). Discarding again the $\mathcal{O}(p^{4})$ terms (which only contribute to the UV-finite part) and performing the $t$-integration of the remaining expression yields:
\begin{eqnarray}
    J_{2}=\frac{\pi^{d/2}\Gamma(\frac{y}{2})}{2\Gamma(\frac{d+y}{2})}\int_{0}^{1}dx \ x^{\frac{d+y-2}{2}}\Bigg[\frac{p_{\parallel}^{2}(\nu+(1-x)\nu_{\bot})^{(1 - d)/2}}{(\nu+(1-x)\nu_{\parallel})^{3/2}} + \nonumber \\
    +\frac{p_{\bot}^{2}(\nu+(1-x)\nu_{\bot})^{(-1 - d)/2}}{(\nu+(1-x)\nu_{\parallel})^{1/2}}\Bigg]f({\bf p})^{\frac{-y}{2}} + \text{finite part}.
\end{eqnarray}
By expanding the last expression in double Laurent series in powers of $\varepsilon$ and $y$ and keeping only negative power term, we get
\begin{eqnarray}
    J_{2}=\frac{\pi^{2}}{2y}\int_{0}^{1}dx \ \Bigg[\frac{p_{\parallel}^{2}(\nu+(1-x)\nu_{\bot})^{-3/2}}{(\nu+(1-x)\nu_{\parallel})^{3/2}} 
    +\frac{p_{\bot}^{2}(\nu+(1-x)\nu_{\bot})^{-5/2}}{(\nu+(1-x)\nu_{\parallel})^{1/2}}\Bigg]
\end{eqnarray}
up to the terms of order $\mathcal{O}(\varepsilon^{0},y^{0})$. The integral part of the last expression exactly coincides with its counterpart in (\ref{I2expanded}), so that the final expression for UV divergent part is
\begin{eqnarray}
    J_{2}=\frac{\pi^{2}}{y}
    \Bigg[\frac{p^{2}_{\parallel}}{\nu^{2}(2\nu + \nu_{\parallel} + \nu_{\bot} + 2\sqrt{(\nu+\nu_{\parallel})(\nu + \nu_{\bot})})} + \nonumber \\
    p^{2}_{\bot}\frac{\nu_{\bot} - 3\nu_{\parallel}-2\nu+2\left(\nu+\nu_{\parallel}
        \sqrt{({\nu+\nu_{\parallel}})/({\nu+\nu_{\bot}})}\right)
    }{3\nu^{2}(\nu_{\parallel}-\nu_{\bot})^{2}}
    \Bigg] + \text{finite part}.
    \label{J2final}
\end{eqnarray}
By substituting (\ref{I1final}), (\ref{I2final}) into (\ref{D1split}) and  (\ref{J1final}), (\ref{J2final}) into (\ref{D2split}), expressing all quantities in terms of charges $x_{1}$, $x_{2}$ and $w$, $u$ rescaled by $8\pi^{2}$, and performing algebraic simplifications, eventually we arrive at the final expression for the UV-divergent part of the diagram:
\begin{eqnarray}
   \hhstrokeSecond = - \frac{w}{ \epsilon} \, 
    \left[ \nu_{\parallel }p^2_\parallel f_1(x_1,x_2) +    \nu_{\perp } p^2_{\perp}f_2(x_1,x_2)  \right] - \nonumber \\ - \frac{u}{ y} \, 
    \left[ \nu_{\parallel }p^2_\parallel f_1(x_1,x_2) +    \nu_{\perp } p^2_{\perp}f_2(x_1,x_2)  \right] 
    \label{fin}
\end{eqnarray}
with the functions $f_i(x_1,x_2)$ defined in (\ref{ff}) and (\ref{fff}). By combining this result with expression (\ref{hktwopoint}), we arrive at the renormalization constants $Z_{1}$ and $Z_{2}$ given in (\ref{Z1}), (\ref{Z2}).

\end{document}

%% file: tikzpics.tex
\def \hhhstrokeone{
\raisebox{-4ex}[0cm][0cm]{\begin{tikzpicture}
\begin{feynman}
\vertex (a1) ;
					\vertex[below=0.4cm of a1] (a2);
					\vertex[below right= 1.5cm of a2] (a3);
					\vertex[below left= 1.5cm of a2] (a4);
				    \vertex[below left=0.4cm of a4] (a5);
				    \vertex[below right=0.4cm of a3] (a6);
					\diagram{
						(a1) --[fermionZerowidth'''] (a2),
						(a2) --[fermion'] (a3),
						(a2) --[fermion'](a4),
						(a5) --[fermion] (a4),
						(a3) --[fermion](a4),
						(a3) --[fermion](a6),
					};
				\end{feynman}

\end{tikzpicture}}
}

\def \hhhstroketwo{
\raisebox{-4ex}[0cm][0cm]{\begin{tikzpicture}
\begin{feynman}
\vertex (a1) ;
					\vertex[below=0.4cm of a1] (a2);
					\vertex[below right= 1.5cm of a2] (a3);
					\vertex[below left= 1.5cm of a2] (a4);
				    \vertex[below left=0.4cm of a4] (a5);
				    \vertex[below right=0.4cm of a3] (a6);
					\diagram{
						(a1) --[fermionZerowidth'''] (a2),
						(a2) --[fermion] (a3),
						(a2) --[fermion'](a4),
						(a4) --[fermion] (a5),
						(a4) --[fermion'](a3),
						(a3) --[fermion](a6),
					};
				\end{feynman}

\end{tikzpicture}}
}

\def \hhhstrokethree{
\raisebox{-4ex}[0cm][0cm]{\begin{tikzpicture}
\begin{feynman}
	\vertex (a1) ;
					\vertex[below=0.4cm of a1] (a2);
					\vertex[below right= 1.5cm of a2] (a3);
					\vertex[below left= 1.5cm of a2] (a4);
				    \vertex[below left=0.4cm of a4] (a5);
				    \vertex[below right=0.4cm of a3] (a6);
					\diagram{
						(a1) --[fermionZerowidth'''] (a2),
						(a2) --[fermion'] (a3),
						(a2) --[fermion](a4),
						(a4) --[fermion] (a5),
						(a3) --[fermion'](a4),
						(a3) --[fermion](a6),
					};
				\end{feynman}
\end{tikzpicture}}
}

\def \hhhstrokefour{
\raisebox{-4ex}[1.7cm][0cm]{\begin{tikzpicture}
\begin{feynman}
							\vertex (a1) ;
					\vertex[below=0.4cm of a1] (a2);
					\vertex[below right= 1.5cm of a2] (a3);
					\vertex[below left= 1.5cm of a2] (a4);
				    \vertex[below left=0.4cm of a4] (a5);
				    \vertex[below right=0.4cm of a3] (a6);
					\diagram{
						(a1) --[fermionZerowidth'''] (a2),
						(a2) --[fermion'] (a3),
						(a2) --[fermion'](a4),
						(a5) --[fermion] (a4),
						(a3) --[Boson](a4),
						(a3) --[fermion](a6),
					};
				\end{feynman}

\end{tikzpicture}}
}

\def \hhhstrokefive{
\raisebox{-4ex}[1.7cm][0cm]{\begin{tikzpicture}
\begin{feynman}
							\vertex (a1) ;
					\vertex[below=0.4cm of a1] (a2);
					\vertex[below right= 1.5cm of a2] (a3);
					\vertex[below left= 1.5cm of a2] (a4);
				    \vertex[below left=0.4cm of a4] (a5);
				    \vertex[below right=0.4cm of a3] (a6);
					\diagram{
						(a1) --[fermionZerowidth'''] (a2),
						(a2) --[fermion'] (a3),
						(a2) --[Boson](a4),
						(a5) --[fermion] (a4),
						(a3) --[fermion'](a4),
						(a3) --[fermion](a6),
					};
				\end{feynman}

\end{tikzpicture}}
}

\def \hhhstrokesix{
\raisebox{-4ex}[1.7cm][0cm]{\begin{tikzpicture}

\begin{feynman}
							\vertex (a1) ;
					\vertex[below=0.4cm of a1] (a2);
					\vertex[below right= 1.5cm of a2] (a3);
					\vertex[below left= 1.5cm of a2] (a4);
				    \vertex[below left=0.4cm of a4] (a5);
				    \vertex[below right=0.4cm of a3] (a6);
					\diagram{
						(a1) --[fermionZerowidth'''] (a2),
						(a2) --[Boson] (a3),
						(a2) --[fermion'](a4),
						(a4) --[fermion] (a5),
						(a4) --[fermion'](a3),
						(a3) --[fermion](a6),
					};
				\end{feynman}
\end{tikzpicture}}
}

\def \hhhstrokeVertex{
    \vcenter{\hbox{
    \begin{tikzpicture}
      \begin{feynman}
\vertex (a1) ;
\vertex[below=0.6cm of a1] (a2);
\vertex[below left=0.6cm of a2] (a5);
\vertex[below right=0.6cm of a2] (a6);
\vertex[above right=0.1cm of a2] (b1) {\({\bf k}\)};
\diagram{
(a1) --[fermionZerowidth'''] (a2),
(a2) --[fermion] (a5),
(a2) --[fermion](a6),
};
        \end{feynman}
    \end{tikzpicture}}} 
}

\def \hvhstrokeVertex{
    \vcenter{\hbox{
    \begin{tikzpicture}
      \begin{feynman}
\vertex (a1) ;
\vertex[below=0.6cm of a1] (a2);
\vertex[below left=0.6cm of a2] (a5);
\vertex[below right=0.6cm of a2] (a6);
\vertex[above right=0.1cm of a2] (b1) {\({\bf k}\)};
\vertex[below left=0.5cm of a2] (b2) {\({\small s}\)};
\diagram{
(a1) --[fermionZerowidth'''] (a2),
(a2) --[Boson] (a5),
(a2) --[fermion](a6),
};
        \end{feynman}
    \end{tikzpicture}}} 
}

\def \vvvstrokeVertex{
    \vcenter{\hbox{
    \begin{tikzpicture}
      \begin{feynman}
\vertex (a1) ;
\vertex[below=0.6cm of a1] (a2);
\vertex[below left=0.6cm of a2] (a5);
\vertex[below right=0.6cm of a2] (a6);
\vertex[above right=0.1cm of a2] (b1) {\({\bf k}\)};
\vertex[below left=0.5cm of a2] (b2) {\({\small s}\)};
\vertex[below right=-0.1cm of a6] (b3) {\({\small n}\)};
\vertex[above=0.6cm of a2] (b4) {\({\small j}\)};
\diagram{
(a1) --[bosonZerowidth'] (a2),
(a2) --[Boson] (a5),
(a2) --[Boson](a6),
};
        \end{feynman}
    \end{tikzpicture}}} 
}

\def \hhstroke{ 
    \vcenter{\hbox{
    \begin{tikzpicture}
      \begin{feynman}
        \vertex (a);
        \vertex [right=1.5cm of a] (b);
    
        \diagram* {
            (a) -- [fermion', thick] (b)
            };
        \end{feynman}
    \end{tikzpicture}}} 
}

\def \vvstrokeHKNS{ 
    \vcenter{\hbox{
    \begin{tikzpicture}
      \begin{feynman}
        \vertex (a);
        \vertex [right=1.5cm of a] (b);
        
        \diagram* {
            (a) -- [boson', thick] (b)
            };
        \end{feynman}
    \end{tikzpicture}}} 
}

\def \lineonly{ 
    \vcenter{\hbox{
    \begin{tikzpicture}
      \begin{feynman}
        \vertex (a);
        \vertex [right=1.5cm of a] (b);
    
        \diagram* {
            (a) -- [fermion, thick] (b)
            };
        \end{feynman}
    \end{tikzpicture}}} 
}

\def \lineonlyboson{ 
    \vcenter{\hbox{
    \begin{tikzpicture}
      \begin{feynman}
        \vertex (a);
        \vertex [right=1.5cm of a] (b);
    
        \diagram* {
            (a) -- [boson, thick] (b)
            };
        \end{feynman}
    \end{tikzpicture}}} 
}

\def \vstrokevvHKNS{ 
    \vcenter{\hbox{
    \begin{tikzpicture}
      \begin{feynman}
        \vertex (a);
        \vertex [above=1cm of a] (b) {\(j\)};
        \vertex [below left=1cm of a] (c) {\(n\)};
        \vertex [below right=1cm of a] (d) {\(s\)};
        \vertex [above right = 0.3 cm of a] {\(k\)};
    
        \diagram* {
            (b) -- [boson''] (a),
            (a) -- [Boson] (c),
            (a) -- [Boson] (d), 
            };
        \end{feynman}
    \end{tikzpicture}}} 
}

\def \hstrokehhHKNS{ 
    \vcenter{\hbox{
    \begin{tikzpicture}
      \begin{feynman}
        \vertex (a);
        \vertex [above=1cm of a] (b);
        \vertex [below left=1cm of a] (c);
        \vertex [below right=1cm of a] (d);
        \vertex [above right = 0.3 cm of a] {\(k\)};
    
        \diagram* {
            (b) -- [SB] (a),
            (a) -- [fermion] (c),
            (a) -- [fermion] (d), 
            };
        \end{feynman}
    \end{tikzpicture}}} 
}

\def \hstrokevhHKNS{ 
    \vcenter{\hbox{
    \begin{tikzpicture}
      \begin{feynman}
        \vertex (a);
        \vertex [above=1cm of a] (b);
        \vertex [below left=1cm of a] (c) {\(i\)};
        \vertex [below right=1cm of a] (d);
        \vertex [above right = 0.3 cm of a] {\(k\)};
    
        \diagram* {
            (b) -- [SB] (a),
            (a) -- [Boson] (c),
            (a) -- [fermion] (d), 
            };
        \end{feynman}
    \end{tikzpicture}}} 
}

\def \hhstrokeFirst{
    \vcenter{\hbox{
    \begin{tikzpicture}
      \begin{feynman}
        \vertex (a);
        \vertex [right=0.8cm of a] (b);
        \vertex [right=of b] (c);
        \vertex [right=0.8cm of c] (e);
    
        \diagram* {
            (a) -- [fermion, thick] (b),
            (c) -- [fermion', half right] (b),
            (b) -- [fermion, half right] (c), 
            (c) -- [fermionZerowidth''] (e),
            };
        \end{feynman}
    \end{tikzpicture}}} 
}
\def \hhstrokeSecond{
    \vcenter{\hbox{
    \begin{tikzpicture}
      \begin{feynman}
        \vertex (a);
        \vertex [right=0.8cm of a] (b);
        \vertex [right=of b] (c);
        \vertex [right=0.8cm of c] (e);
    
        \diagram* {
            (a) -- [fermion] (b),
            (c) -- [fermion', half right] (b),
            (b) -- [->, boson, half right, thick] (c), 
            (c) -- [fermionZerowidth''] (e),
            };
        \end{feynman}
    \end{tikzpicture}}} 
}

\def \vvstroke{
    \vcenter{\hbox{
    \begin{tikzpicture}
      \begin{feynman}
        \vertex (a);
        \vertex [right=0.8cm of a] (b);
        \vertex [right=of b] (c);
        \vertex [right=0.8cm of c] (e);

        \diagram* {
            (a) -- [Boson] (b),
            (c) -- [boson', half right] (b),
            (b) -- [Boson, half right, looseness=1.5] (c), 
            (c) -- [bosonZerowidth'] (e),
            };
        \end{feynman}
    \end{tikzpicture}}} 
}

\def \vvstrokeNSdetail{ 
    \vcenter{\hbox{
    \begin{tikzpicture}
      \begin{feynman}
        \vertex (a);
        \vertex [right=0.8cm of a] (b);
        \vertex [right=of b] (c);
        \vertex [right=0.8cm of c] (e);a
        \vertex [above=0.4cm of b, red] (b1) {\(l\)};
        \vertex [below=0.4cm of b, red] (b2) {\(s\)};
        \vertex [above=0.4cm of c, red] (c1) {\(l'\)};
        \vertex [below=0.3cm of c, red] (c2) {\(s'\)};
    
        \diagram* {
            (a) -- [Boson, edge label'={\(i\)}, momentum={[arrow style=blue]\(p\)}] (b),
            (b) -- [antiboson', half left, reversed momentum'={[arrow shorten=0.25]\(\omega\)}, momentum={[arrow shorten=0.35, arrow style=blue]\(p-k\)}] (c),
            (b) -- [Boson, half right, looseness=1.5, momentum'={[arrow shorten=0.35, arrow style=blue]\(k\)}] (c), 
            (c) -- [bosonZerowidth', edge label'=\(j\), momentum={[arrow style=blue]\(p\)}] (e),
            };
        \end{feynman}
    \end{tikzpicture}}} 
}

\def \hhstrokeFirstdetail{ 
    \vcenter{\hbox{
    \begin{tikzpicture}
      \begin{feynman}
        \vertex (a);
        \vertex [right=0.8cm of a] (b);
        \vertex [right=of b] (c);
        \vertex [right=0.8cm of c] (e);
    
        \diagram* {
            (a) -- [fermion, momentum={[arrow style=blue]\(p\)}] (b),
            (b) -- [antifermion', half left, reversed momentum'={[arrow shorten=0.25]\(\omega\)}, momentum={[arrow shorten=0.35, arrow style=blue]\(p-k\)}] (c),
            (b) -- [fermion, half right, looseness=1.5, momentum'={[arrow shorten=0.35, arrow style=blue]\(k\)}] (c), 
            (c) -- [fermionZerowidth'', momentum={[arrow style=blue]\(p\)}] (e),
            };
        \end{feynman}
    \end{tikzpicture}}} 
}

\def \hhstrokeSeconddetail{ 
    \vcenter{\hbox{
    \begin{tikzpicture}
      \begin{feynman}
        \vertex (a);
        \vertex [right=0.8cm of a] (b);
        \vertex [right=of b] (c);
        \vertex [right=0.8cm of c] (e);
        \vertex [below=0.4cm of b, black] (b2) {\(a\)};
        \vertex [below=0.3cm of c, black] (c2) {\(b\)};
    
        \diagram* {

            (a) -- [fermion, momentum={\(p\)}] (b),
            (b) -- [antifermion', half left, reversed momentum'={[arrow shorten=0.25]\(\omega\)}, momentum={[arrow shorten=0.35]\(p-k\)}] (c),
            (b) -- [Boson, half right, looseness=1.5, momentum'={[arrow shorten=0.35]\(k\)}] (c), 
            (c) -- [fermionZerowidth'', momentum={\(p\)}] (e),
            };
        \end{feynman}
    \end{tikzpicture}}} 
}

\tikzfeynmanset{ SB/.style = { 
   decoration={markings,
        mark=at position 0.6 with {\arrow[semithick]{Stealth[black,width=3mm,length=3mm]}},
        mark=at position 0.8 with {\arrow[semithick]{Bar[black,width=3mm,length=0mm]};}},
     line width=0.3mm, 
   postaction=decorate}
}

\tikzfeynmanset{ fermion'/.style = { 
   decoration={
     markings,
     mark=at position 0.7
          with {\arrow[xshift=2mm]{Bar[black,width=3mm,length=0mm]}}
     },
     line width=0.3mm, 
   postaction=decorate}
}

\tikzfeynmanset{ fermion/.style = {
     line width=0.3mm,
   postaction=decorate}
}

\tikzfeynmanset{ boson'/.style = {boson, 
   decoration={
     markings,
     mark=at position 0.7
          with {\arrow[xshift=2mm]{Bar[black,width=3mm,length=0mm]}}
     },
     line width=0.3mm, 
   postaction=decorate}
}

\tikzfeynmanset{ antiboson'/.style = {boson, 
   decoration={
     markings,
     mark=at position 0.2
          with {\arrow[xshift=0mm]{Bar[black,width=3mm,length=0mm]}}
     },
     line width=0.3mm, 
   postaction=decorate}
}

\tikzfeynmanset{ bosonZerowidth'/.style = {boson, 
   decoration={
     markings,
     mark=at position 0.05
          with {\arrow[xshift=2mm]{Bar[black,width=3mm,length=0mm]}}
     },
     line width=0.3mm, 
   postaction=decorate}
}

\tikzfeynmanset{ antifermion'/.style = { 
   decoration={
     markings,
     mark=at position 0.2
          with {\arrow[xshift=0mm]{Bar[black,width=3mm,length=0mm]}}
     },
     line width=0.3mm, 
   postaction=decorate}
}

\tikzfeynmanset{ Boson/.style = {boson, 
     line width=0.3mm,
   postaction=decorate}
}

\tikzfeynmanset{ fermionZerowidth''/.style = { 
   decoration={
     markings,
     mark=at position 0.05
          with {\arrow[xshift=2mm]{Bar[black,width=3mm,length=0mm]}}
     },
     line width=0.3mm, 
   postaction=decorate}
}

\tikzfeynmanset{ fermionZerowidth'''/.style = { 
   decoration={
     markings,
     mark=at position 0.05
          with {\arrow[xshift=2.5mm]{Bar[black,width=3mm,length=0mm]}}
     },
     line width=0.3mm, 
   postaction=decorate}
}

\tikzfeynmanset{ boson''/.style = {boson, 
   decoration={markings,
        mark=at position 0.6 with {\arrow[semithick]{Stealth[black,width=3mm,length=3mm]}},
        mark=at position 0.8 with {\arrow[semithick]{Bar[black,width=3mm,length=0mm]};}},
     line width=0.3mm, 
   postaction=decorate}
}